\pdfoutput=1
\documentclass[a4paper,fleqn,usenatbib]{mnras}

\usepackage[T1]{fontenc}
\usepackage{ae,aecompl}
\usepackage{graphicx}
\usepackage{physics}
\usepackage{amsmath}

\usepackage{graphicx}	
\usepackage{amsmath}	
\usepackage{amssymb}	
\newcommand{\msun}{{\rm M}_\odot}
\usepackage{newtxtext,newtxmath}

\title{Neutrino emission from the collapse of $\sim 10^4$ $M_\odot$ population III supermassive stars }

\author[C. Nagele et al.]{
Chris Nagele,$^{1}$\thanks{E-mail: chrisnagele.astro@gmail.com}
Hideyuki Umeda,$^{1}$
Koh Takahashi, $^{2}$
Takashi Yoshida,$^{3}$
Kohsuke Sumiyoshi$^{4}$
\\
$^{1}$Department of Astronomy, Graduate School of Science, the University of Tokyo, Tokyo, 113-0033, Japan\\
$^{2}$Max Plank Institute for Gravitational Physics (Albert Einstein Institute), D-14476 Potsdam, Germany\\
$^{3}$Yukawa Institute for Theoretical Physics, Kyoto University, Kyoto 606-8502 Japan\\
$^{4}$National Institute of Technology, Numazu College, Ooka 3600, Numazu, Shizuoka 410-8501, Japan
}

\date{Accepted XXX. Received YYY; in original form ZZZ}

\pubyear{2020}

\begin{document}
\label{firstpage}
\pagerange{\pageref{firstpage}--\pageref{lastpage}}
\maketitle

\begin{abstract}
We calculate the neutrino signal from Population III supermassive star collapse using a neutrino transfer code originally developed for core collapse supernovae and massive star collapse. Using this code, we are able to investigate the supermassive star mass range thought to undergo neutrino trapping ($\sim 10^4$ $\msun$), a mass range which has been neglected by previous works because of the difficulty of neutrino transfer. For models in this mass range, we observe a neutrino-sphere with a large radius and low density compared to typical massive star neutrino-spheres. We calculate the neutrino light-curve emitted from this neutrino-sphere. The resulting neutrino luminosity is significantly lower than the results of a previous analytical model. We briefly discuss the possibility of detecting a neutrino burst from a supermassive star or the neutrino background from many supermassive stars and conclude that the former is unlikely with current technology, unless the SMS collapse is located as close as 1 Mpc, while the latter is also unlikely even under very generous assumptions. However, the supermassive star neutrino background is still of interest as it may serve as a source of noise in proposed dark matter direct detection experiments. 
\end{abstract}

\begin{keywords}
stars: Population III -- gravitation -- stars: black holes -- neutrinos
\end{keywords}



\section{Introduction}
\label{introduction}

In recent years, the continued discovery of high redshift quasars \citep{mortlock2011,wu2015,banados2018,matsuoka2019,wang2021} has lead to an increasing awareness and interest in the early universe supermassive black hole (SMBH) problem, specifically, the question of how SMBHs were created so early in the universe. \citet{rees1984} laid out a number of possibilities for AGN formation and one of the most promising of those in the early universe is the direct collapse scenario. Direct collapse refers to a primordial massive halo which does not cool below the atomic cooling threshold \citep{latif2015,hirano2017} and instead collapses into a supermassive star (SMS) \cite{bromm2003}. Depending on its mass \citep{fuller1986,montero2012,chen2014,nagele2020} and accretion rate \citep{hosokawa2012,hosokawa2013,schleicher2013,umeda2016,woods2017,hammerle2018a,woods2021}, the SMS evolves until it becomes unstable to radial perturbations in general relativity (GR) \citep{chandrasekhar1964,fuller1986,haemmerle2020}. At this point, the SMS contracts and one of three outcomes occurs. The SMS is stabilized by further nuclear burning and continues to evolve, the SMS explodes due to rapid alpha capture burning \citep{chen2014,nagele2020,moriya2021}, or the SMS collapses to a black hole \citep[e.g.][]{shapiro1979,liu2007}. During this collapse, the SMS may emit neutrinos, gravitational waves \citep{shibata2016,uchida2017,li2018,hartwig2018} and an ultra long gamma ray burst (ULGRB) \citep{gendre2013,matsumoto2015,sun2017}.

In the vast majority of cases, the SMS does not explode and it collapses to form a black hole. Several works have investigated SMS collapse and the ensuing neutrino light-curve \citep{woosley1986,shi1998,linke2001,montero2012}. All of these studies found that the collapse of a zero metallicity SMS is smooth and initially homologous. Since the SMS is supported primarily by radiation pressure, the density is low enough that degeneracy pressure is negligible before the final second of the collapse. Because of this lack of degeneracy, as the central temperature approaches $1.1$ MeV, electron positron pairs are created freely. The abundance of charged leptons and the high temperature leads to neutrino cooling via electron/positron capture and charged lepton pair annihilation. This cooling will cause the collapse to become non homologous \citep{fuller1986,shi1998}. As the temperature increases further, the neutrino energy increases and neutrino trapping becomes relevant. Specifically, neutrino scattering on charged leptons which are still abundant due to the lack of degeneracy (and to a lesser extent on nucleons) and nucleon absorption of neutrinos causes the neutrinos to become trapped within a neutrino-sphere which has a large radius and low density compared to that of typical massive stars \citep[][]{sumiyoshi2007}.

Previous studies of collapsing SMS neutrino light-curves can be split into two categories. First, \citet{shi1998} used analytic methods to calculate the neutrino luminosity from a collapsing SMS as a function of the mass of the homologous core ($M^{\rm HC}$). They found that the luminosity scales as $(M^{\rm HC})^{-1.5}$, so that lower mass stars which reach higher final temperatures (at apparent horizon formation) will have stronger neutrino emission. In their approach, they assumed that the neutrino emission came primarily from pair neutrinos and that no neutrino trapping was present.

The second category consists of simulations of high mass SMSs ($M > 10^5$ $\msun$) which are not hot enough for neutrino trapping to occur, thus greatly simplifying the calculation \citep{woosley1986,linke2001,montero2012}. \citet{woosley1986} used the results of \citet{fuller1986} and the neutrino transfer code of \citet{bowers1982} to calculate the neutrino light-curve of a $M = 5 \times 10^5$ $\msun$ collapsing SMS. They found that it was optically thin to neutrinos. \citet{linke2001} (see also \citet{montero2012}) used a 1D GR code with local neutrino emission to calculate the light-curves of collapsing SMSs in the mass range $M =  10^{5-8}$ $\msun$. Like \citet{shi1998}, they found that neutrino luminosity decreases with increasing mass. The goal of this paper is to start to fill in the gap between $\sim 100$ $\msun$ Pop III stars and $M > 10^5$ $\msun$ Pop III supermassive stars, a mass range which has not been investigated thus far because of the difficulty of neutrino transfer simulations.

If SMSs were plentiful in the early universe and produced a large number of neutrinos when collapsing to black holes, then it is possible that the neutrino background (sometimes referred to as relic neutrinos) may have a contribution from SMS neutrinos \citep[e.g.][]{woosley1986}. The neutrino background has two components which are both of physical interest and theoretically detectable (for an overview, see \citealt{vitagliano2020}). First, the cosmic neutrino background (C$\nu$B) is the neutrino equivalent of the cosmic microwave background \citep{lesgourgues2013,desalas2017}. This consists of neutrinos which existed at the time of neutrinos decoupling from matter, about one second after the big bang \citep{vitagliano2020}. A straightforward calculation shows that these neutrinos should have very low energy $\sim 10^{-4}$ eV. Attempts to detect C$\nu$B using the KATRIN and MARE experiments are underway \citep[e.g.][]{hodak2011}, while the PTOLEMY experiment is being specifically designed for this purpose \citep{betti2019}.

Another component of the neutrino background is neutrinos from core collapse supernovae (CCSN) and massive star collapse \citep{nakazato2015}. This component has much higher energy ($\sim 10$ MeV) than the C$\nu$B, and an upper bound on its density may be derived from galactic metallicity constraints \citep[e.g.][]{totani1996}. Several ongoing neutrino projects aim to detect this CCSN background including Super-Kamiokande GD (SK Gd), Hyper-Kamiokande, JUNO, BOREXINO, and KamLAND \citep[e.g.][]{vitagliano2020}.

The SMS neutrino background would be similar to the supernova neutrino background, though with lower energy and likely a lower neutrino density \citep{munoz2021}. We briefly discuss the detectability of the SMS neutrino background under basic assumptions about the density of SMSs in the early universe. We find that even under the most optimistic of these assumptions, the SMS neutrino background is detectable neither by Cherenkov detectors such as SK nor by liquid scintilator detectors such as KamLAND. \citet{munoz2021} reached a similar conclusion regarding coherent nucleon neutrino scattering in proposed dark matter detection experiments. They note that although SMS neutrinos may not be directly detectable in these experiments, they could serve as a source of noise to dark matter direct detection.

In Sec. \ref{methods}, we describe the three codes used in this paper. In Sec. \ref{progenitor}, we detail the results of the stellar evolution calculation and the progenitor models for the collapse. We describe this collapse in Sec. \ref{char} and the resulting neutrino light-curve in Sec. \ref{nsphere}. Finally, in Sec. \ref{detection}, we discuss prospects for detection. We conclude with a discussion in Sec. \ref{discussion}.

\section{Methods}
\label{methods}

\subsection{Stellar Evolution}
\label{HOSHI}
In this paper, we use the same stellar evolution code (HOSHI) as in \citet{nagele2020}, which included a 49 isotope nuclear network, neutrino cooling, and the first order post Newtonian correction to GR. We also adopt the switching condition from \citet{nagele2020}, specifically, we switch from HOSHI to the hydrodynamics calculation when $ dt_{\rm HOSHI} \ll  \tau_{\rm KH}$ where $dt_{\rm HOSHI}$ is the time step in HOSHI and $\tau_{\rm KH}$ is the Kelvin-Helmholtz timescale. For lower mass models, ($M\leq 3\times 10^4$ $\msun$), which are more stable against general relativity, this switching condition is satisfied at higher temperatures than the higher mass models. For the lower mass models, we proceed directly to the neutrino transfer code. 

Note that $m_{\rm r}$ refers to the mass inside a given radius $r$, $\rho$ is the density, $s$ the entropy per baryon, $T$ the temperature. $P$ the pressure, and $X(i)$ the mass fraction of a nuclear species $i$. Quantities with a subscript c such as $\rho_{\rm c}$ refer to the central value of that quantity.

\subsection{Hydrodynamics}
\label{HYDnuc}

After the HOSHI calculation, we switch to a hydrodynamics code with nuclear reactions (HYDnuc) \citep{yamada1997,takahashi2016}, which was also used in \citet{nagele2020}. The HYDnuc code is based on a 1+1D spherical, Lagrangian, general relativistic, neutrino radiation hydrodynamics code (nuRADHYD) developed by \citet{yamada1997} and modified in \citet{sumiyoshi2005}. HOSHI and HYDnuc have the same EOS for low temperature \citep{takahashi2016}, the same nuclear network and the same neutrino cooling reactions \citep{itoh1996}. Thus, the primary differences between HYDnuc and HOSHI are as follows: a) HYDnuc has full GR, though at low densities the post Newtonian correction in HOSHI should be a good approximation, b) HYDnuc includes an acceleration term where HOSHI does not, and c) HYDnuc does not treat energy transport beyond cooling via neutrino emission, while HOSHI includes convective energy transport according to 1D mixing length theory. These differences are important when discussing whether a SMS will explode, but in the case of black hole collapse, they are not as relevant.

\subsection{Hydrodynamics with neutrino transfer}
\label{nuRADYD}

Around $T = 0.85$ MeV, the electron capture and positron capture reactions start to have a significant impact on the evolution of the electron fraction, $Y_e = (n_p - n_n)/N_A$ because of the abundance of free electrons and positrons produced via pair production (here, $n_p,n_n$ are the number densities of neutrons and protons and $N_A$ is the Avogadro constant). Note that the definition of $Y_e$ does not include $e^-,e^+$ pairs, but the neutrino reaction rates do account for $e^-,e^+$ pairs. At $T = 0.2-0.8$ MeV, we switch to the neutrino transfer version of HYDnuc, nuRADHYD \citep{yamada1997,sumiyoshi2005}. 

\begin{figure}
    \centering
    \includegraphics[width=\columnwidth]{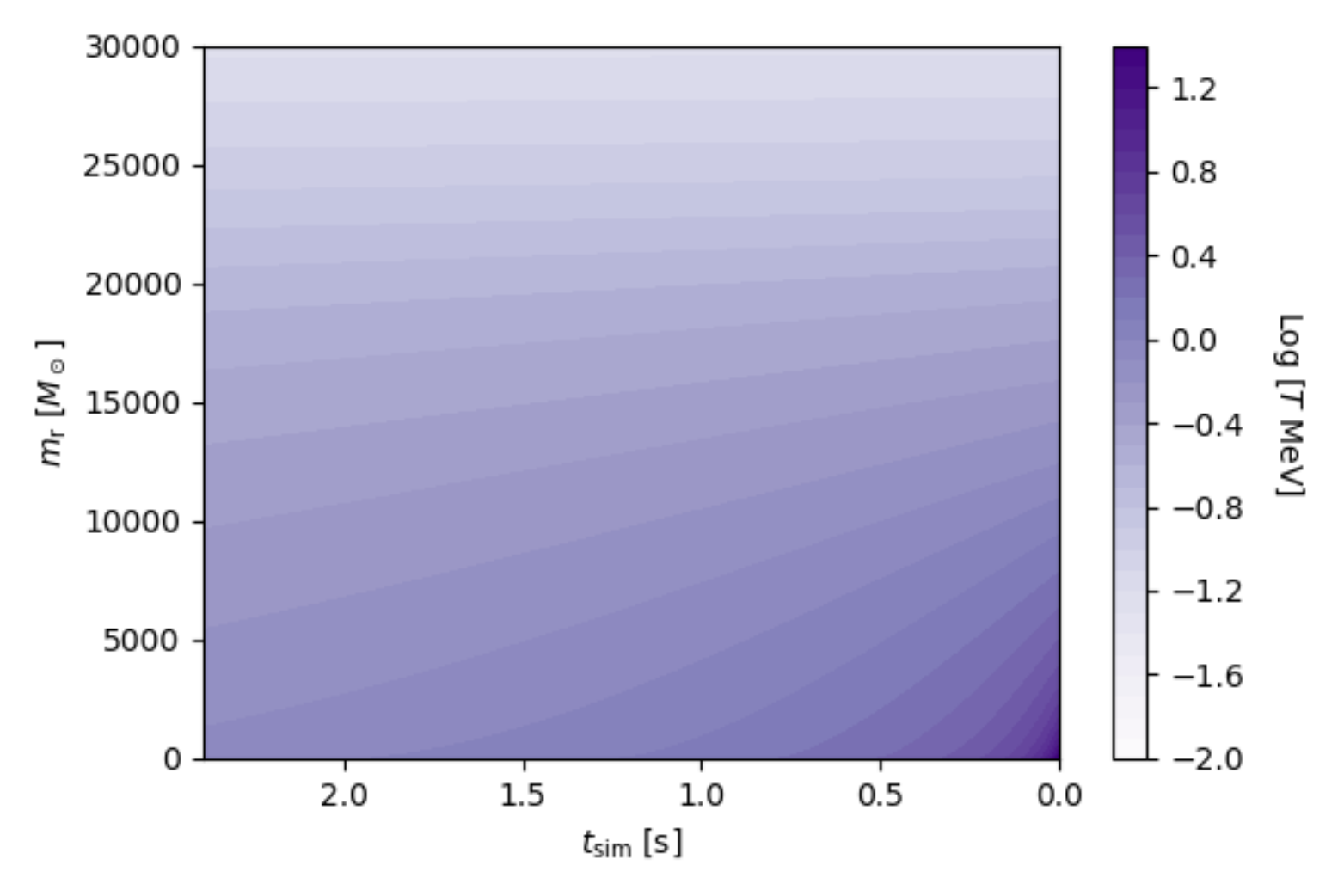}
	
    \caption[Temperature evolution in nuRADHYD.]{Temperature evolution of the $M=5.5\times 10^4$ $\rm  M_\odot$ model (subsequent figures will also use this model unless otherwise specified) in nuRADHYD. Despite significant nuclear burning, weak reactions, and photodissociation, the increase in temperature is monotonic and smooth due to the dominant role of gravity. $t_{\rm sim}$ is time until the end of the simulation. }
    \label{fig:Tdens}
\end{figure}

\begin{figure}
    \centering
    \includegraphics[width=\columnwidth]{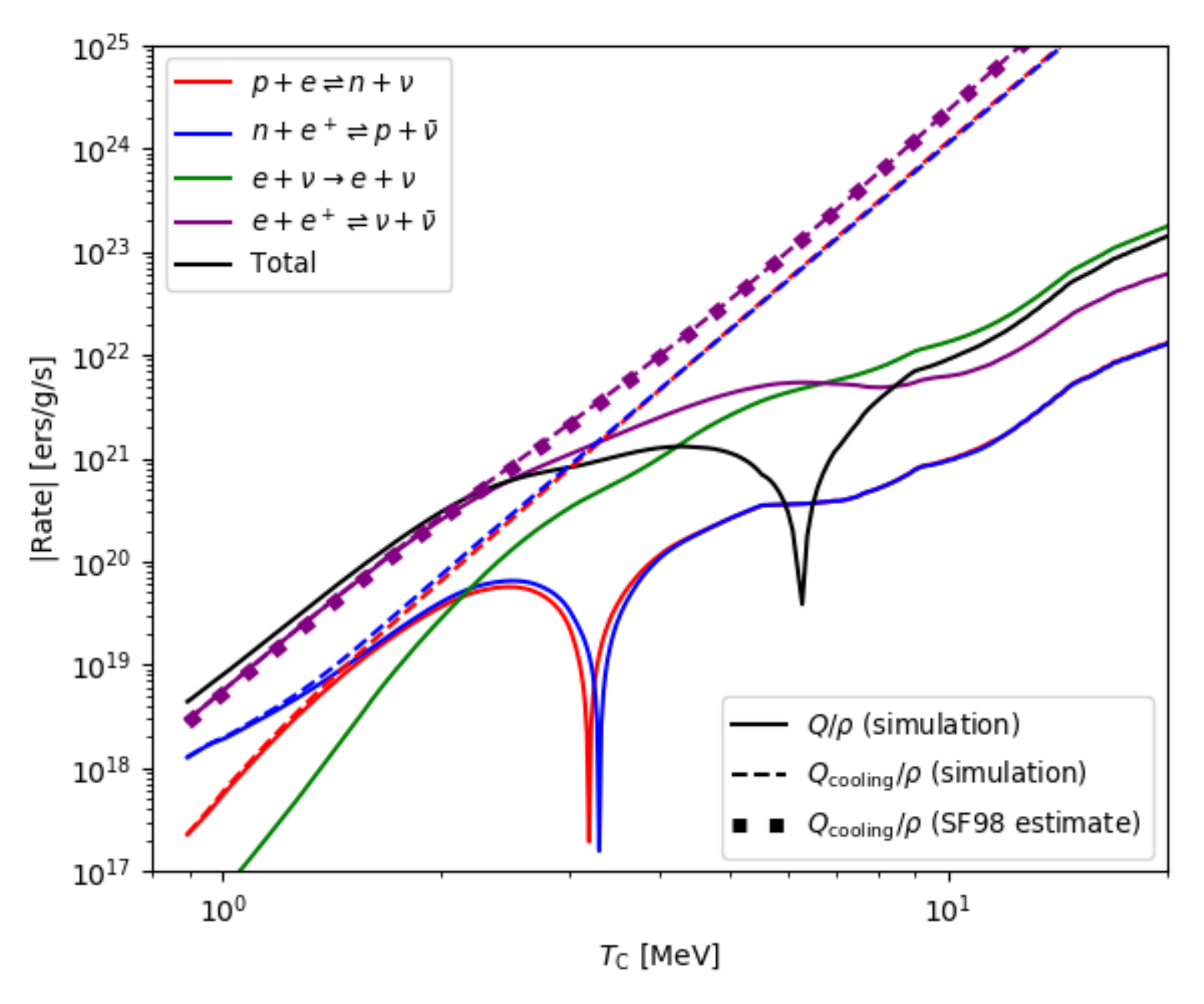}
	
    \caption[Absolute value of the neutrino emissivity $Q$ in nuRADHYD for various neutrino reactions as a function of central temperature.]{Absolute value of central neutrino emissivity $Q$ in nuRADHYD for various neutrino reactions as a function of central temperature. $Q$ [ergs/g/s] is the total emissivity and $Q_{\rm cooling}$ is this quantity only from neutrino emission. Initially, neutrino reactions cool the star $Q < 0$, but around $T_{\rm c} \approx 6$ MeV, neutrino reactions start to heat the star, roughly corresponding with the minimum of entropy in Fig. \ref{fig:Mrs}. This heating is related to neutrino thermalization (Appendix B). Pair emissivity (purple) is compared to the analytic estimate of \citet{shi1998} and good agreement is found. Note that for $T_{\rm c} > 2 $ MeV, $Q_{\rm cooling} \gg Q$ so in this regime, both cooling and heating reactions are necessary.}
    \label{fig:emiss}
\end{figure}

The temperature increases monotonically and smoothly in space (inwards) and time (Fig. \ref{fig:Tdens}). At $T \approx 2$ MeV, the neutrino flux is large enough that nucleon absorption of neutrinos becomes important (Fig. \ref{fig:emiss}). Thus, for SMSs with $T_{\rm c}>2$ MeV, it is necessary to include neutrino transfer.

The Misner Sharp metric \citep{misner1964} can be written in natural units as:
\begin{equation}
    ds^2 = e^{2\phi} dt^2 - e^{2 \lambda} dm_{\rm r}^2 - r^2 d\Omega^2
\end{equation}
where $\phi$ and $\lambda$ are metric components which vary with $t,m_{\rm r}$, and $d\Omega^2 = d\theta^2 +\sin^2 \theta d\phi^2$. Then, we can define the gamma factor $\Gamma$ and the gravitational mass $\Tilde{m_{\rm r}}$ using the constraint equations from \citet{yamada1997}:
\begin{equation}
    \pdv{\Tilde{m_{\rm r}}}{m_{\rm r}}= 4 \pi r^2 \bigg[\rho (1 + \epsilon + \epsilon_\nu)+\frac{vF_\nu}{\Gamma}\bigg] \pdv{r}{m_{\rm r}}
    \label{eq:gmass}
\end{equation}
\begin{equation}
    \Gamma^2 = 1 + \rm{v}^2  - \frac{2\Tilde{m_{\rm r}}}{r}.
\end{equation}
where $\rm{v}$ is the velocity, $\epsilon_\nu$ and $F_\nu$ are the neutrino energy density and flux vector respectively, and $\epsilon$ is the internal energy density.

The gravitational mass will be noticeably larger than the baryonic mass when the internal energy is high or the neutrino energy is high. The condition for the apparent horizon may be written as either of the following \citep{sumiyoshi2007}:
\begin{equation}
    \rm{v} + \Gamma \leq 0
\end{equation}
\begin{equation}
    r = 2 \Tilde{m_{\rm r}}.
    \label{eq:apparent}
\end{equation}
The code terminates within $\sim 0.02$ s of apparent horizon formation when various physical quantities give convergence problems and the time-step decreases drastically. The major limitations of nuRADHYD in this context relate to the Misner Sharp metric, specifically the fact that it only considers one spatial dimension and cannot handle the black hole singularity.

Several modifications were made to nuRADHYD in order to accommodate the high temperature, low density environment of SMS collapse. First, unlike the CCSN scenario, in SMS collapse, protonization occurs (see Sec. \ref{char}) so we expect $Y_e > 0.5$. The original Shen EOS only covers $Y_e \leq 0.5$ \citep{shen1998} so the EOS table was extended to $Y_e = 0.65$ using the modified Shen EOS \citep{shen2011}. However, this is not sufficient because $Y_{e, \rm max} \approx 0.7$ in the collapsing SMS models. In the high $Y_e$ region ($Y_e > 0.6$), we calculate the chemical potentials of protons and neutrons using the Boltzmann chemical potential. This is a valid approximation because the electron fraction is high, so the situation must be non degenerate. We then set $X_p = Y_e$, $X_n = 1- Y_e$ in the high $Y_e$ region.
    
The next change was to include accurate expressions for the rest mass contribution to the chemical potentials of nucleons. In the CCSN case, the chemical potentials of protons and neutrons are greater than 938 MeV such that
\begin{equation}
    \mu_n - 938 \;[\rm MeV]\; \gg \Delta 
\end{equation}
where $\Delta=m_n - m_p$. The approximation $m_n = m_p =938$ MeV (relativistic mean field theory) holds for CCSN \citep{sumiyoshi2005}, but not for the case of a collapsing SMS, so we include accurate expressions for the rest mass in order to get realistic weak reaction rates. 

\begin{figure*}
    \centering
    \includegraphics[width=2\columnwidth]{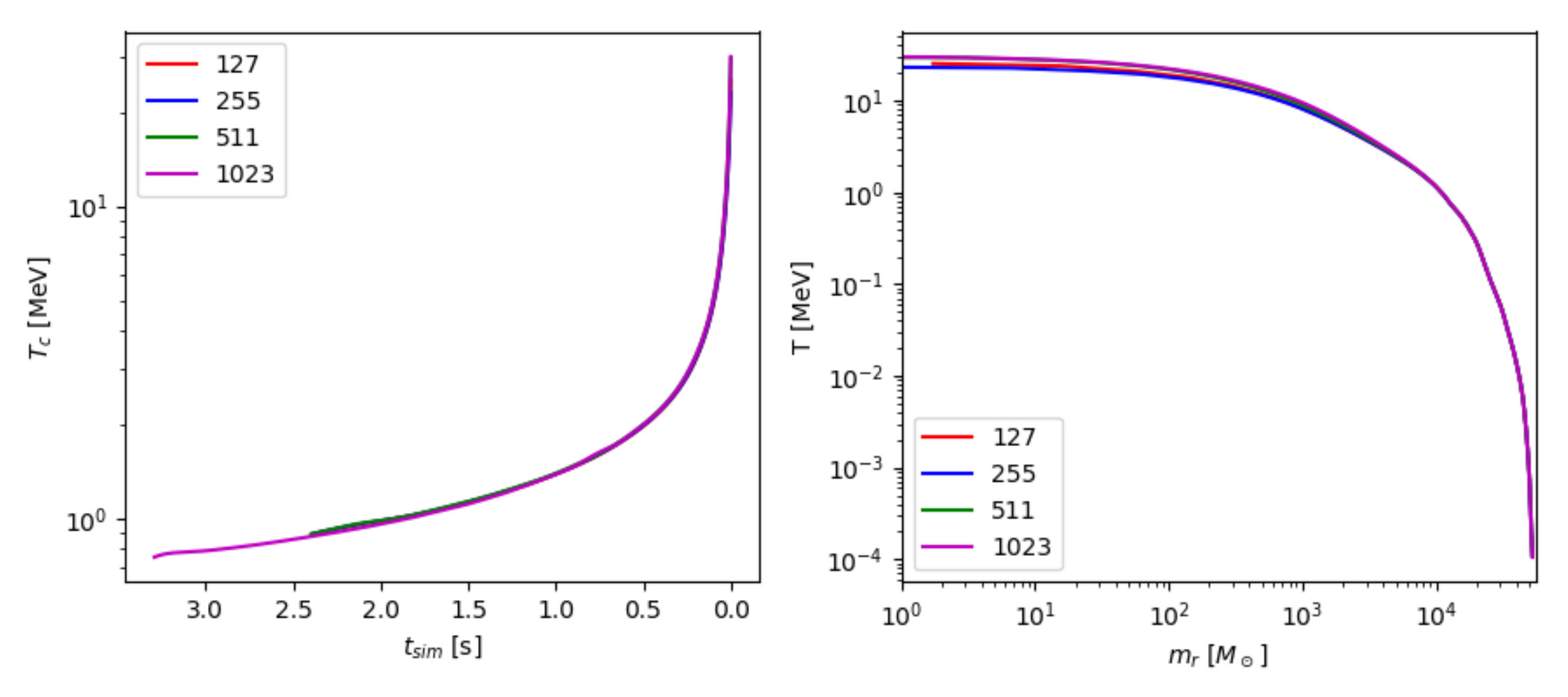}

    \caption{Numerical convergence plots for four different values of the radial mesh number in nuRADHYD. Left panel - time evolution of central temperature. Right panel - temperature profile at final time step. Note that the initial data for 1023 is different from the initial data for the others because we had to enlarge the mesh number in HYDnuc to accommodate 1023 meshes in nuRADHYD. In both figures good agreement is found for all four values of the radial mesh number. For the rest of the paper we use 511 radial meshes. }
    \label{fig:nconv}
\end{figure*}

Similarly to switching from HOSHI to HYDnuc, the major change when switching from HYDnuc to nuRADHYD is the redefinition of the radial mesh. We find excellent agreement for several different radial resolutions (Fig. \ref{fig:nconv}) and we use 511 radial meshes for all models in this paper. The time when switching from HYDnuc to nuRADHYD does not affect our results as long as the central temperature is below 1 MeV.

\section{Results}
\label{results}

\subsection{Collapse progenitors}
\label{progenitor}

We consider 11 SMS models (Table \ref{tab:models}) ranging in mass from $10^4$ $\msun$ to $6 \times 10^4$ $\msun$. Models with mass above $5 \times 10^4$ $\msun$ are taken from \citet{nagele2020} while the lower mass models were made in the same way as in \citet{nagele2020}, but they are new to this paper. For some of the low mass models ($1-3 \times 10^4$ $\msun$), we use a 52 isotope nuclear network instead of 49 isotopes with the additional three isotopes being $^{14}$O, $^{18}$Ne, and $^{19}$Ne \citep{takahashi2019}. The models with the 52 isotope nuclear network are more stable against the GR instability during hydrogen burning. The change in isotope number does not greatly effect other physical parameters. SMSs with mass lower than $10^4$ $\msun$ begin to resemble Pop III massive stars. SMSs with mass higher than $6 \times 10^4$ $\msun$ are of interest, particularly with regards to determining the maximum mass with a neutrino-sphere present and to allow direct comparisons to previous works which studied higher mass models. However, because of the difficulty of evaluating the stability of these high mass models during the stellar evolution calculation, we leave this to future work. 

\begin{figure}

    \includegraphics[width=\columnwidth]{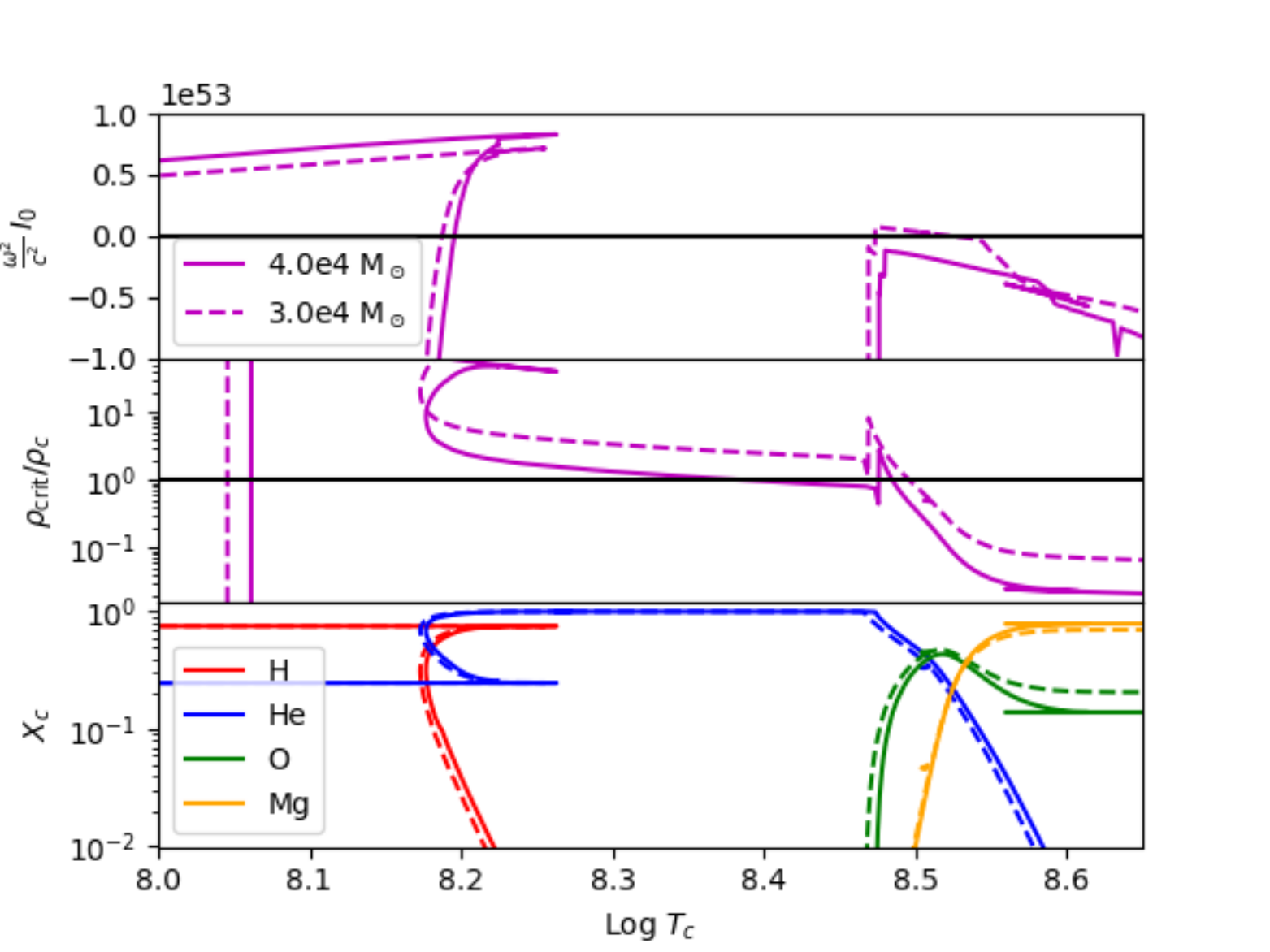}
    \caption{Several quantities showing the stability of the SMS models in HOSHI. Upper panel - Stability of radial perturbation in general relativity using Eq. 11 of \citet{haemmerle2020} for two models, $M=3.0, 4.0 \times 10^4$ $\msun$. The vertical axies shows $\omega^2/c^2 I_0$ and if this value is positive, the model is stable and vice versa. Middle panel - $\rho_{\rm crit}/\rho_c$ from \citet{fuller1986}. If the vertical axis is greater than unity, the model is GR stable and vice versa. Lower panel - Chemical mass fractions for the central mesh. Note that the $4.0 \times 10^4$ $\msun$ model is the 52 isotope model, not the 49 isotope model used later in this paper. We do this simply to get a better comparison between the chemical mass fraction in the lower panel. Other physical parameters are identical between the 49 and 52 isotope models and the number of isotopes does not effect the upper two panels of this plot.   }
    \label{fig:GRinst}
\end{figure}

In Fig. \ref{fig:GRinst}, the $3.0\times 10^4$ $\msun$ model is an example of one of the low mass, more stable against GR models, while the $4.0\times 10^4$ $\msun$ model is representative of the higher mass, less stable models. In order to explain this difference, we analyze the stability of these two models using two GR instability criteria, that of \citet{fuller1986} (polytropic criterion) and of \citet{haemmerle2020}, which are both derived from a variational analysis of radial perturbations \citep{chandrasekhar1964}. Both approaches evaluate Eq. 61 of \citet{chandrasekhar1964}, where \citet{fuller1986} evaluates the integrals in Eq 61. analytically by assuming a polytropic EOS and \citet{haemmerle2020} assumes a specific form of the trial function, then carries out integration by parts on Eq 61. followed by numerical integration over the star.

The \citet{haemmerle2020} criterion (Fig. \ref{fig:GRinst} upper panel) shows that both stars are unstable until the start of helium burning. Then the low mass star becomes stable while the high mass star remains unstable, thus explaining why the low mass star reaches a higher temperature before collapse. On the other hand, the \citet{fuller1986} criterion (Fig. \ref{fig:GRinst} middle panel) shows that both stars are stable until the higher mass star becomes unstable during helium burning, thus explaining why the high mass star collapses earlier. The key difference in these results is that the polytropic criterion does not view the star as unstable during the Kelvin Helmholtz contraction between hydrogen and helium burning (Log $T_{\rm c} \approx 8.3 - 8.4$), while the criterion of \citet{haemmerle2020} does. 

It is also important to note that neither of these criteria correspond to the star becoming dynamical, which occurs at Log $T_{\rm c}$= 8.8 for the $4.0 \times 10^4$ $\msun$ model and  Log $T_{\rm c}$= 8.7 for the $4.0 \times 10^4$ $\msun$ model, and the reason is that neither criterion takes into account the effect of future nuclear burning, which can stabilize the star against collapse. Thus the lower mass models continue nuclear burning to higher temperatures because of this difference in stability. By the time the lower mass SMSs enter dynamical collapse, they have completely burnt the core helium and have started production of silicon. In comparison, the higher mass models still have helium present in the core at dynamical collapse. Indeed, it is this helium which facilitated the explosions reported in \citet{chen2014} and \citet{nagele2020}.

\subsection{Characteristics of collapse}
\label{char}

\begin{figure}

    \centering
	\includegraphics[width=\columnwidth]{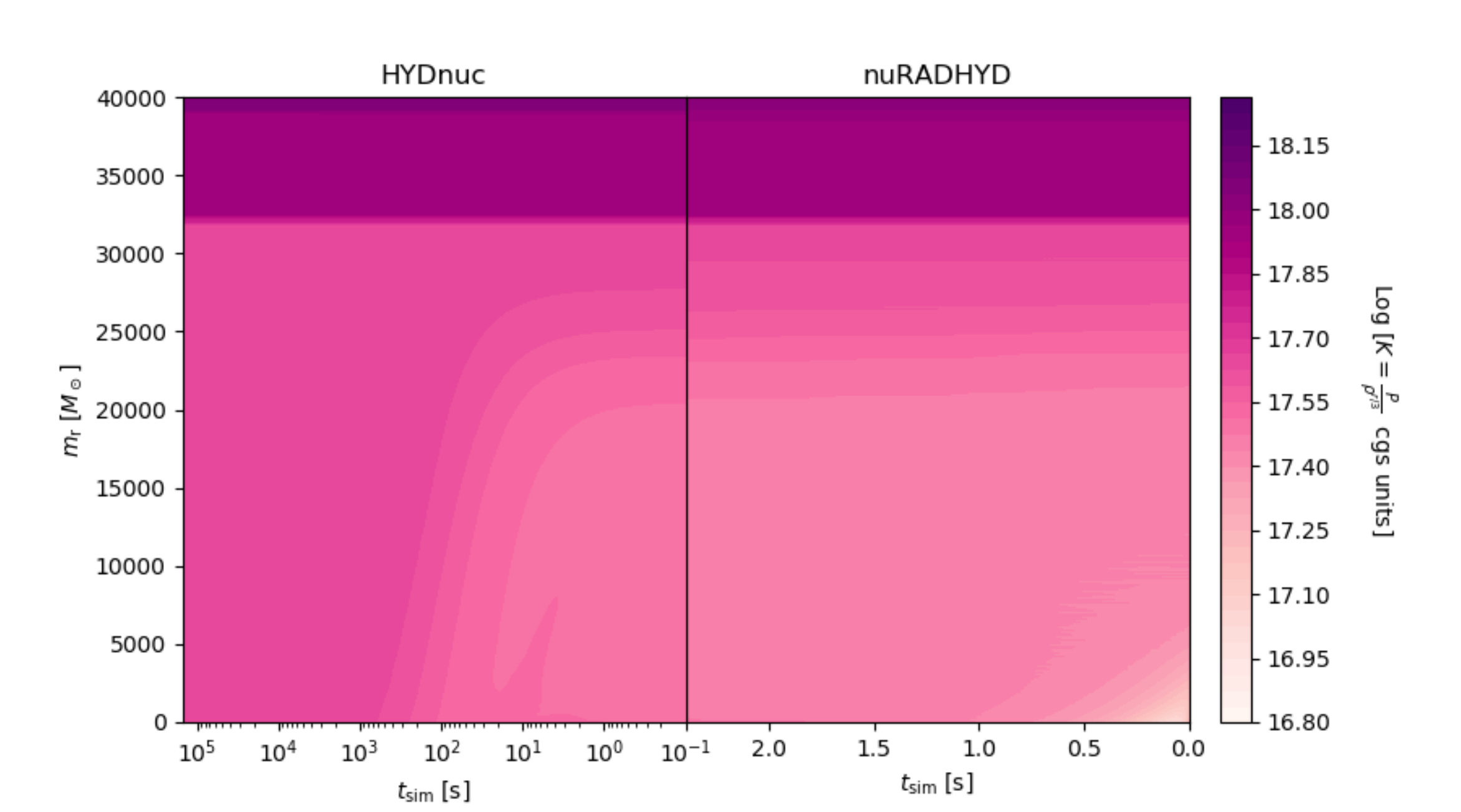}
    \caption[Polytropic structure in HYDnuc and nuRADHYD.]{Polytropic structure in HYDnuc and nuRADHYD. At the beginning of the HYDnuc calculation, the entire isentropic core is an $n=3$ polytrope. Towards the end of HYDnuc, heavy isotopes photodissociate into helium and the polytropic structure breaks down. Eventually, the inner core again resembles an n=3 polytrope, but this relation breaks down due to strong GR towards the end of nuRADHYD. When switching from HYDnuc to nuRADHYD, the radial meshes are redefined which is why the contours do not exactly line up between the two codes. }
    \label{fig:profiles}
\end{figure}

Previous works \citep{shi1998} assumed that the collapse would proceed homologously until weak reactions start to cool the core (around 1 MeV), and this seems a reasonable assumption for the inner core. \citet{shi1998} also assumed an $n=3$ polytrope structure instead of calculating progenitors using a stellar evolution code as in Sec. \ref{progenitor}. Similar to the homologous assumption, we find that the $n=3$ polytrope assumption is initially true (at least in the core), but it breaks down when heavy isotopes photo-dissociate into helium (Fig. \ref{fig:profiles}). Finally, \citet{shi1998} also assumed $\rho_c \propto T_c ^3$ and this assumption also breaks down due to GR. We note that even though the collapse is proceeding quickly (the in-fall velocity is an appreciable fraction of the speed of light), radiation pressure is large and it keeps the acceleration mostly below Newtonian free fall.

\begin{figure*}
    \centering
    \includegraphics[width=2\columnwidth]{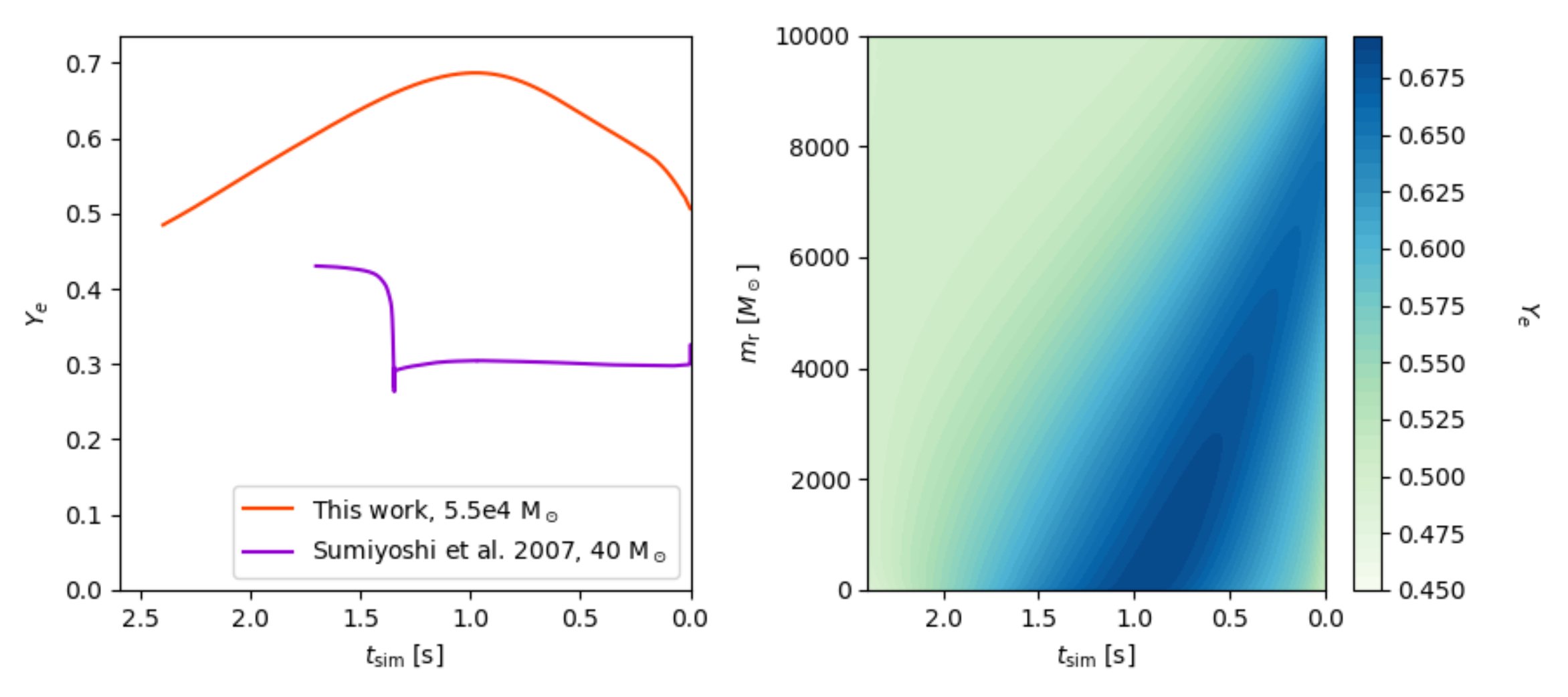}

    \caption[Time evolution of central $Y_e$ in nuRADHYD compared to a $40$ $\rm  M_\odot$ model from \citet{sumiyoshi2007}.]{Left panel - Time evolution of central $Y_e$ in nuRADHYD compared to a $40$ $\rm  M_\odot$ model from \citet{sumiyoshi2007}. The SMS first experiences protonization, then neutrinosiation as degeneracy becomes important. The $40$ $\rm  M_\odot$ model experiences neutronization during the burst. Right panel: Distribution of $Y_e$. Note that at the final neutrino-sphere mass coordinate ($\approx 7000$ $\rm  M_\odot$, Fig. \ref{fig:profiles}), $Y_e$ is increasing for most of the simulation. }
    \label{fig:Ye}
\end{figure*}

At $T\approx 0.45$ MeV, heavy isotopes photo-disassociate into $^4\rm He$ and at $T\approx 0.65$ MeV, the $^4\rm He$ photo-disassociates into nucleons. In the nucleonic region, the electron mass fraction ($Y_e$) evolves according to nucleonic weak reactions (Fig. \ref{fig:Ye}). When the $e^-,e^+$ pair production threshold is reached, the density is still low enough that electron and positron pairs are created freely. This means that both electron capture and positron capture can proceed. However, since positron capture is kinematically preferred ($m_n > m_p$), the fraction of protons will increase (protonization). This is opposite to the neutronization in CCSN, the primary difference being the lack of degeneracy pressure in the low density SMS. As the proton mass fraction increases, $Y_e$ will also increase (Fig. \ref{fig:Ye}). This continues until degeneracy starts to become important ($\rho \sim 10^7$ g/cc), at which point $Y_e$ returns to 0.5. It is worth noting the time scale of protonization is on the order of seconds, which is much longer than the neutronization burst of both CCSN and massive star collapse (Fig. \ref{fig:Ye}).

\begin{figure*}

    \includegraphics[width=2\columnwidth]{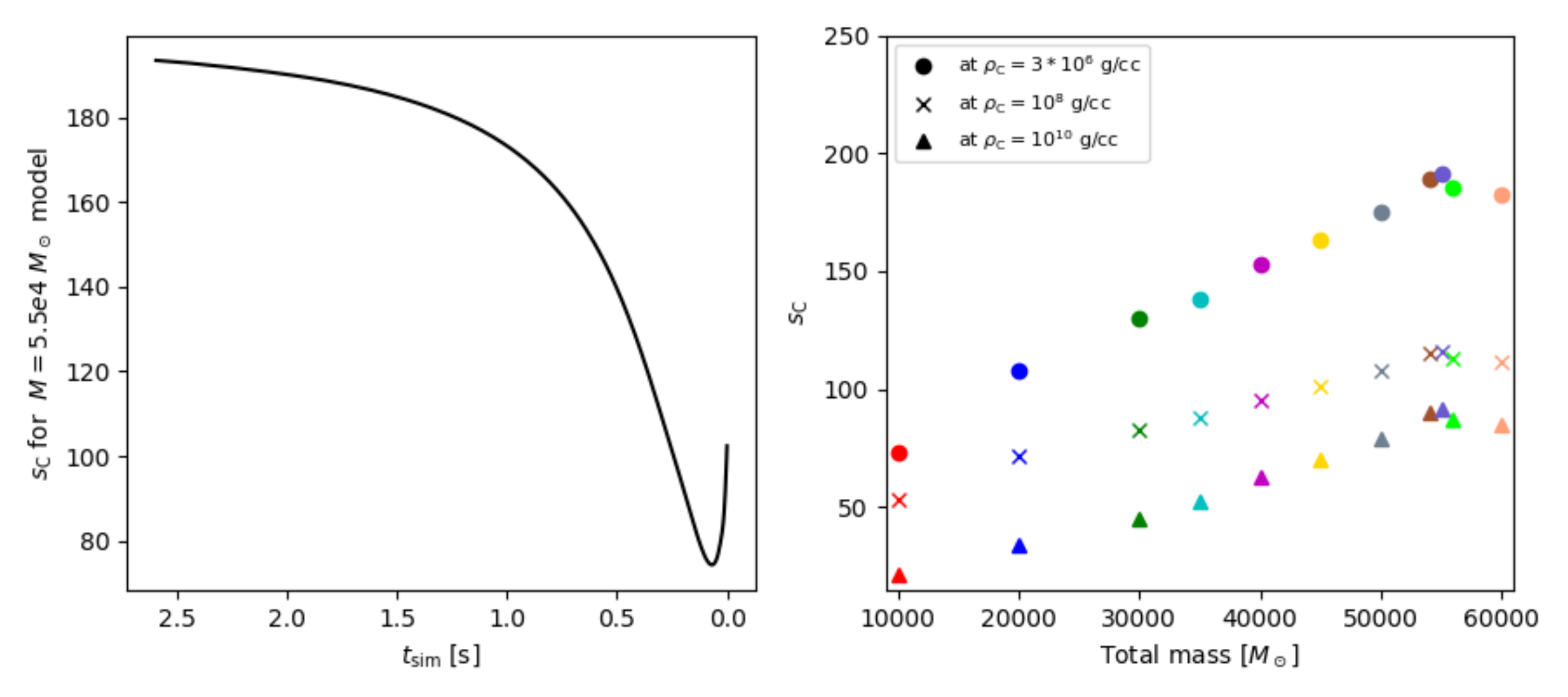}
    \caption{Left panel - Central entropy time evolution in nuRADHYD for the $5.5 \times 10^4$ $\msun$ model. The entropy first decreases due to neutrino cooling and then increases from neutrino heating (Appendix B). Right panel - Central entropy as a function of mass for several densities during the nuRADHYD calculation.}
    \label{fig:Mrs}
\end{figure*}

In the nucleonic region, no nuclear reactions occur and change in entropy is determined solely by neutrino reactions. At the start of nuRADHYD, pair neutrinos and electron/positron capture neutrinos decrease the central entropy (Fig. \ref{fig:Mrs} - Left panel), but at $T \approx 6$ MeV, electron scattering and pair annihilation reactions start to heat the gas as neutrinos are thermalized. This thermalization occurs separately for e-type neutrinos and for x-type neutrinos and is discussed in more detail in Appendix B. The heating from neutrino thermalization increases the entropy and this event roughly corresponds to the change in sign of neutrino emissivity (Fig. \ref{fig:emiss}). All models follow this pattern and for any given density, the hierarchy of entropies remains constant (Fig. \ref{fig:Mrs} - Right panel). Note that models with $M\leq 3.0\times 10^4$ $\msun$ have slightly increased entropy because of prolonged nuclear burning before collapse. 

\begin{figure*}

    \includegraphics[width=2.0\columnwidth]{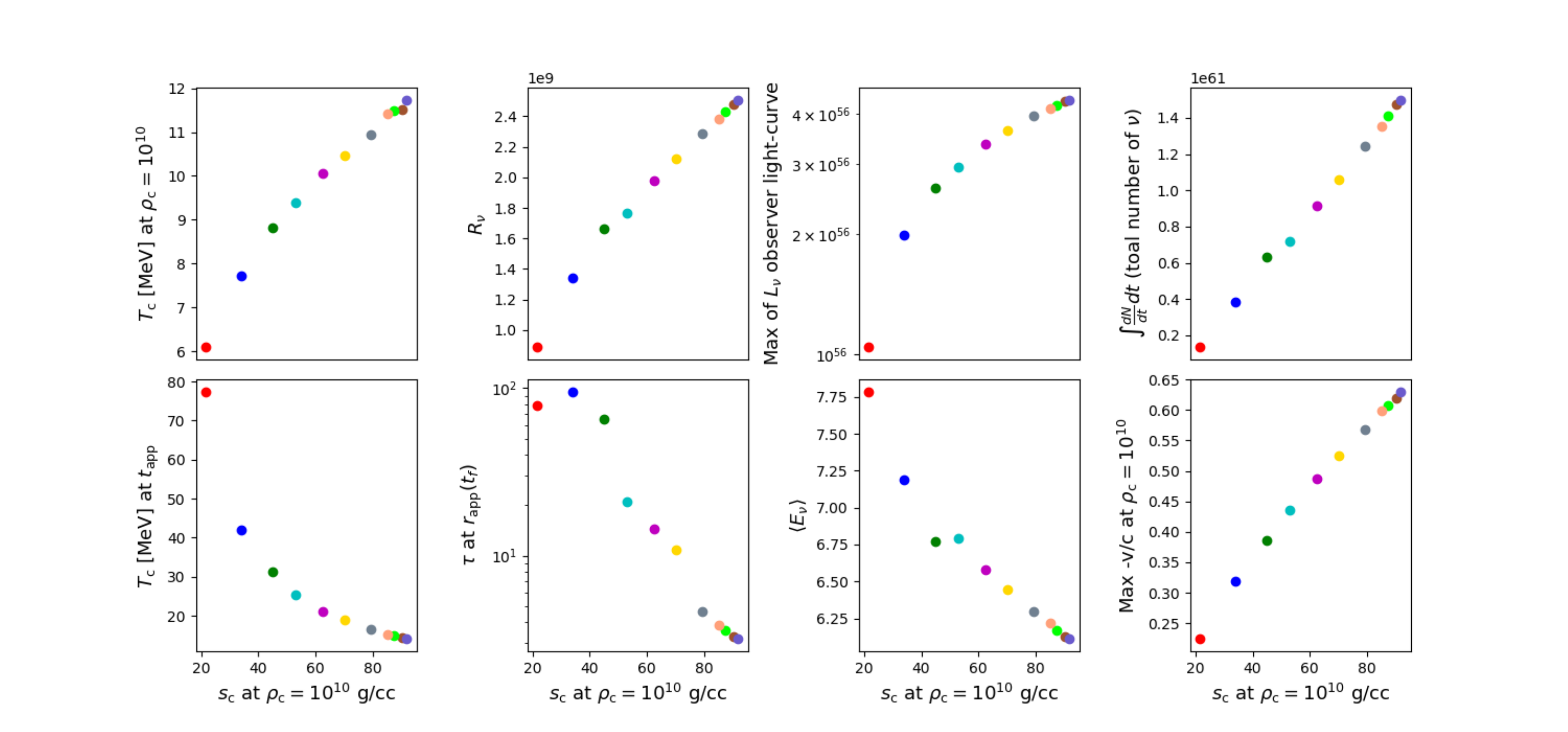}
    \caption{Various quantities as a function of central entropy for a given density. Colors correspond to those in Fig. \ref{fig:Mrs}. First upper panel - Central temperature at fixed density. First lower panel - Central temperature at the apparent horizon formation time-step. Second upper panel- Electron type neutrino-sphere radius. Second lower panel - Optical depth of electron neutrinos at the apparent horizon during the final time-step. Third upper panel - Maximum electron neutrino luminosity for a distant observer. Third lower panel - Integrated average energy. Fourth upper panel - Total electron neutrino number. Fourth lower panel - Maximum infall velocity at fixed density.  }
    \label{fig:scomp}
\end{figure*}

We find that all physical quantities of our simulations scale with the central entropy (Fig. \ref{fig:scomp}) because of the dominant role of gravity. Because our models do not all have the same amount of nuclear burning during their evolution, entropy does not increase monotonically with mass (Fig. \ref{fig:Mrs}). At the beginning of nuRADHYD, $e^{2\phi} \approx 0.7$. This large deviation from unity demonstrates how important a role gravity plays in the collapsing SMSs. The time dilation caused by the extreme curvature is readily apparent in Fig. \ref{fig:Lt}. The apparent horizon forms soon before the simulation terminates and the typical mass is $10^{2-3}$ $\rm  M_\odot$ which is $\sim 1\%$ of the total stellar mass (Fig. \ref{fig:scomp} upper right panel). We will denote the initial apparent horizon time as $t_{\rm app}$ and the outermost trapped surface radius as $r_{\rm app}(t)$. At $r_{\rm app}(t_{\rm app})$, the energy density of matter and neutrinos is roughly equal so both matter and neutrinos contribute to the horizon formation beyond the rest mass contribution. Previous works have only estimated the matter contribution as they assumed free streaming of neutrinos in the higher mass regime.

We will now describe the panels of Fig. \ref{fig:scomp}. All of the neutrino quantities are for electron type neutrinos, but x-type neutrinos undergo similar trends (e.g. Table \ref{tab:models}). The first upper panel shows the central temperature at fixed density, while the first lower panel shows the central temperature at the apparent horizon formation time-step $t_{app}$, and these two quantities have opposite trends with central entropy. For higher entropies, the black hole forms earlier in the collapse because of greater energy density, so that the collapse occurs at lower temperature. This is the primary reason for the decreasing neutrino luminosity as a function of mass for the high mass models considered by previous works \citep{shi1998,linke2001}. However, for the models in this work, the apparent horizon is inside the neutrino-sphere as demonstrated by the second lower panel (see Sec. \ref{nsphere}). Indeed, the neutrino-sphere radius in the second upper panel increases as a function of entropy, though there are two different trends for the different core types (Table \ref{tab:models}). 

Because the neutrino-sphere radius and temperature both increase as a function of entropy, it is reasonable that the neutrino luminosity should also increase, and this can be seen in the third upper panel. Note that for higher mass models, for instance those considered in \citet{shi1998} and \citet{linke2001}, the opposite scaling relation with temperature is found because the temperature at the apparent horizon formation, which determines the neutrino luminosity if no trapping is present, decreases with entropy (first lower panel). We discuss these differences further in Sec. \ref{discussion}. In the third lower panel, the average neutrino energy (integrated over time and neutrino number) decreases with entropy. A large part of this decrease is explained by the increasing gravitational redshift felt by the neutrinos, while the rest is due to the lower neutrino-sphere temperature for larger $R_\nu$. In the fourth upper panel, we have the total neutrino number while in the fourth lower panel, we have the maximum in-fall velocity. Once again we stress that the mass and entropy hierarchies are different so the linear trend in the velocity figure is a general relativistic phenomenon. Note that for the quantities at fixed density, we have chosen $\rho_c = 10^{10}$ g/cc as a rough order of magnitude comparison to $\rho(t_{\rm app})$. Changing this density will change the values in the figures, but not the overall trends as the entropy hierarchy remain fixed throughout the hydrodynamics calculations (Fig. \ref{fig:Mrs}).

\subsection{Neutrino-sphere and light-curve}
\label{nsphere}

SMS collapse may leave observables such as ULGRBs, gravitational waves and neutrinos. Since our code has only one spatial dimension, we focus on neutrinos. In order to determine the neutrino light-curve, we would ideally like to measure the neutrino luminosity at the surface of the SMS. Unfortunately this is not possible because the surface is $10^5$ light seconds away from the core, where neutrino emission primarily occurs and our code is not stable for long after apparent horizon formation. Because we are unable to measure the neutrino luminosity at the surface, we measure it at the neutrino-sphere and apply a gravitational redshift to the resulting light-curve.

Collapsing SMSs have high temperature and low density compared to CCSN and massive star collapse. The low density suggests that neutrinos should freely escape from the core and previous works assumed that this was the case for very high mass SMSs ($\sim 10^5$ M$\odot$) \citep{shi1998,linke2001}. Although the SMS in this work ($\sim 10^4$ M$_\odot$) still have low density compared to massive stars, the high temperature and consequent high neutrino energy causes a neutrino-sphere to form (Figs. \ref{fig:mfp}, \ref{fig:scomp}). This neutrino-sphere is several orders of magnitude larger than the typical CCSN neutrino-sphere. In Appendix A we calculate the neutrino-sphere using analytic estimates depending on $T,\; \rho$ and $\langle E_\nu \rangle$ and Fig. \ref{fig:mfp} shows a comparison of this calculation to the result obtained from the simulation. In the rest of this paper, we use the neutrino-sphere obtained from the simulation.

\begin{figure}

    \includegraphics[width=\columnwidth]{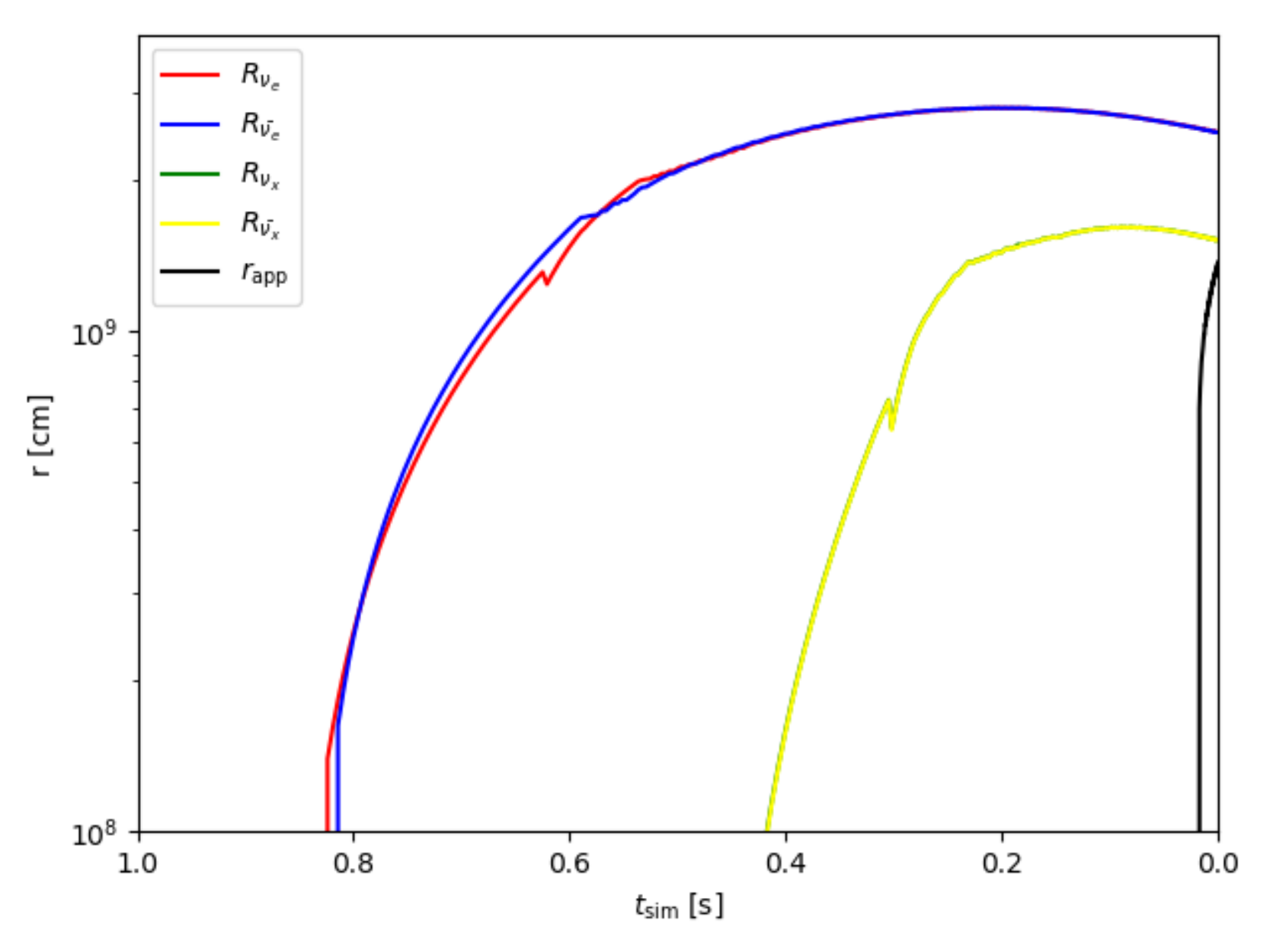}
    \caption{ Time evolution of neutrino-sphere with average energy for each neutrino species. Also shown is the outermost trapped surface, $r_{\rm app}(t)$. }
    \label{fig:nspheret}
\end{figure}

\begin{figure*}

    \includegraphics[width=2\columnwidth]{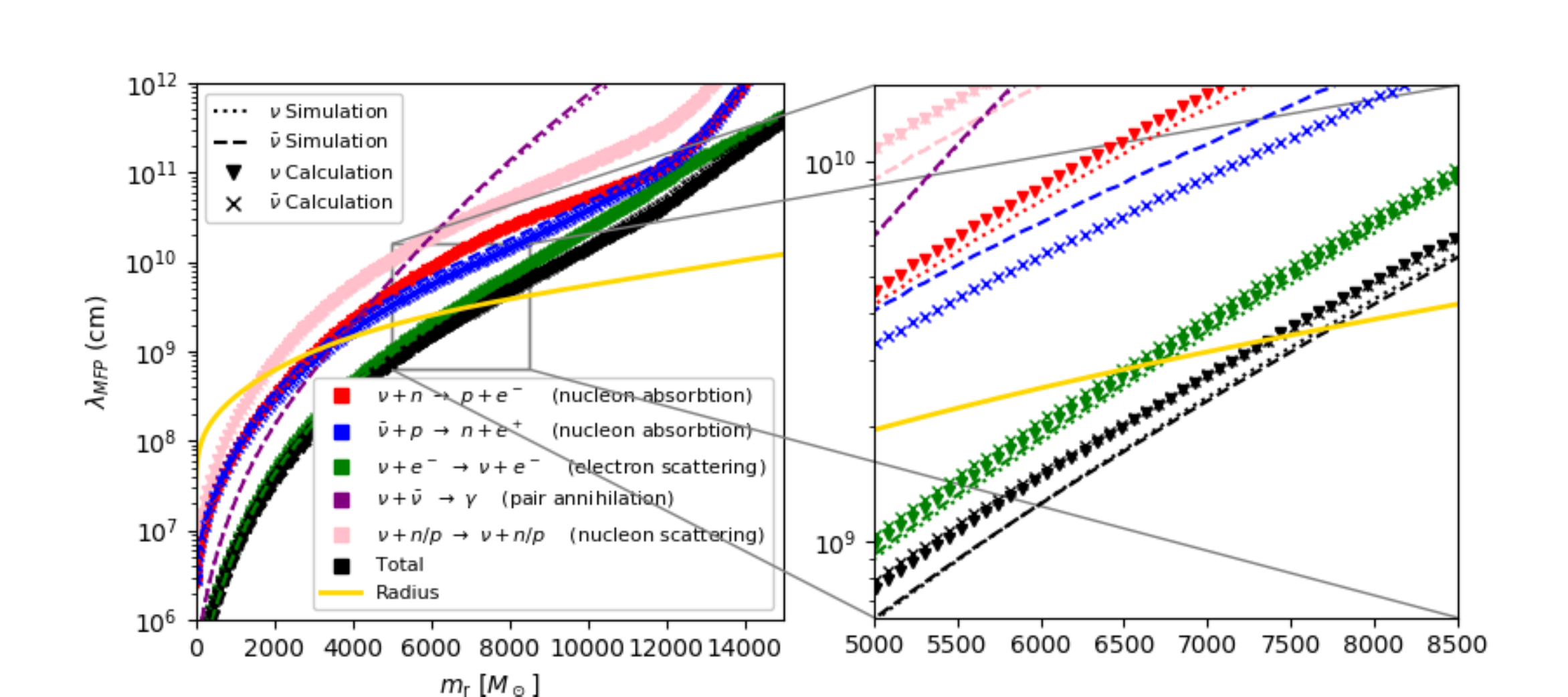}
    \caption[Neutrino mean free paths for $\nu,\bar{\nu}$ from nuRADHYD and calculated using cross sections.]{Left panel: Neutrino mean free paths for $\nu,\bar{\nu}$ from nuRADHYD (lines) and calculated using cross sections (symbols). The neutrino-sphere for each species is located at the intersection of the radius (gold line) with the mean free path. Right panel: Zoomed in to show the neutrino-sphere. }
    \label{fig:mfp}
\end{figure*}

\begin{figure*}
    \centering
    \includegraphics[width=2.0\columnwidth]{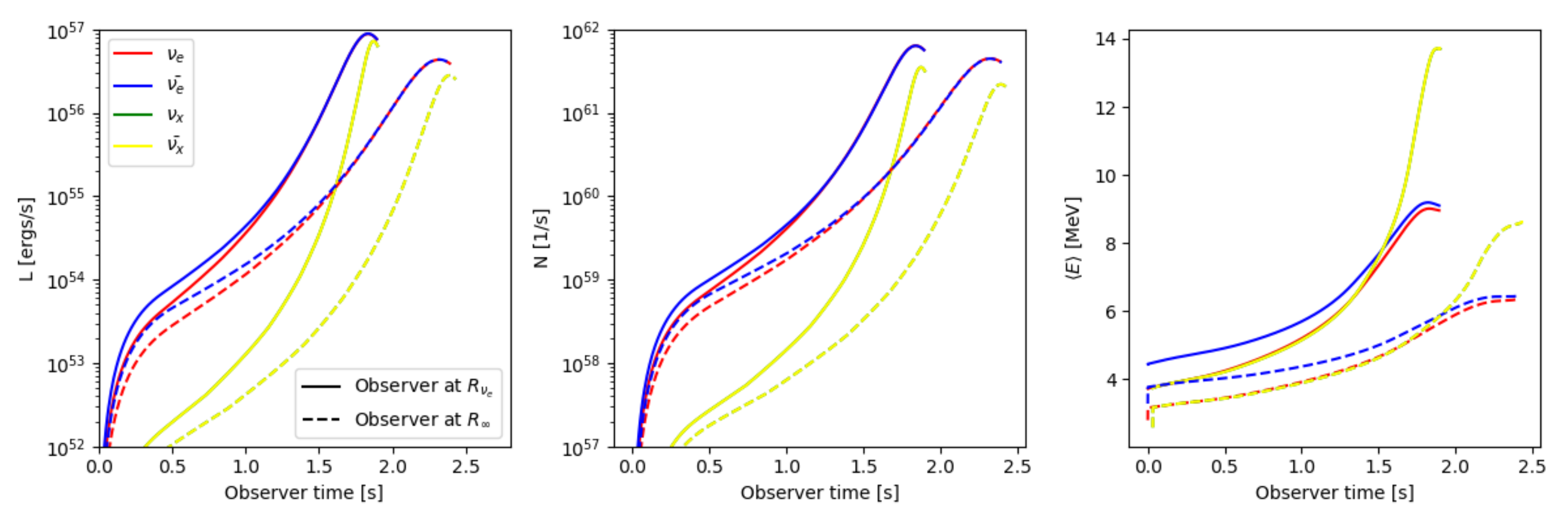}
	
    \caption[Neutrino light-curves for all four species for a local observer at the electron type neutrino-sphere and for a distant observer.]{Left panel: Neutrino light-curves for all four species for a local observer at the electron type neutrino-sphere (solid) and for a distant observer (dashed). The distant observer's light-curve is more spread out due to time dilation. The differences in magnitude between local and distant observers are due to gravitational redshift of neutrinos and decrease in number luminosity from  time dilation. Central panel: Same as left but for total number luminosity. Right panel: Same as left but for average neutrino energy.  }
    \label{fig:Lt}
\end{figure*}

Once we obtain the location of this neutrino-sphere, we can calculate the neutrino light-curve. At the neutrino-sphere radius for each species, the neutrino luminosity, total number luminosity, and average energy of that species is recorded. The solid lines in Fig. \ref{fig:Lt} show the light-curve for a local observer at the $\nu_{e}$ neutrino-sphere radius. Note that the neutrino-spheres for different species are distinct, so the travel time from e.g. $R_{\nu_{x}}$ to $R_{\nu_{e}}$ must be included. $R_{\nu_{x}}$ is in general deeper in the core than $R_{\nu_{e}}$ (Fig. \ref{fig:nspheret}). Thus because the code is terminated at the same time for all meshes, the light-curve of $\nu_{x}$ will extend farther in time than that of $\nu_{e}$.

In order to determine the neutrino light-curve for local and distant observers, the gravitational redshift and time dilation must be taken into account \citep[e.g. Sec. III of ][]{burrows1986}. The gravitational redshift is accounted for using the time-time component of the metric, 
\begin{equation}
    E_{\infty} = E_{\rm Sim} e^{\phi(t,m)}
\end{equation}
where the subscripts $\infty$ and Sim denote a distant observer and the simulation value at the neutrino-sphere radius, respectively. $\phi(t,m)$ is calculated at the neutrino-sphere for each neutrino flavor. The light-curve of the distant observer is the dashed line in Fig. \ref{fig:Lt}.

For the local $R_{\nu_{e}}$ observer, time dilation must be taken into account. The local neutrino-sphere light-curve will spread due to time dilation as it travels out of the star. Similarly to above, the time-time component of the metric is utilized, 
\begin{equation}
    \Delta t_{R_\nu} = \Delta t_{\rm Sim} e^{\phi(t,m)}
\end{equation}
where the $\Delta t_{R_\nu}$ refers to the local observer time, shown by the solid line in Fig. \ref{fig:Lt}, and $\Delta t_{\rm Sim}$ refers to the simulation time corresponding to that of a distant observer, shown by the dashed lines. In Fig. \ref{fig:Lt}, both the solid and dashed lines have been arbitrarily normalized to start at $t=0$. Finally, in order to conserve lepton number, the number luminosity of neutrinos must also be changed by the time dilation factor. 
\begin{equation}
    N_{R_\nu} = N_{\rm Sim} e^{-\phi(t,m)}
\end{equation}
so that the total number of neutrinos, 
\begin{equation}
    \int_0^\infty Ndt
\end{equation}
is the same for both observers. Thus, the neutrino luminosity becomes 
\begin{equation}
    L_{\infty} = L_{\rm Sim} e^{\phi(t,m)}
\end{equation}
\begin{equation}
    L_{R_\nu} = L_{\rm Sim} e^{-\phi(t,m)}
\end{equation}
so that the difference between the luminosities for the two observers is $e^{2\phi(t,m)}$ (Fig. \ref{fig:Lt}).

\begin{figure}
    \centering
    \includegraphics[width=\columnwidth]{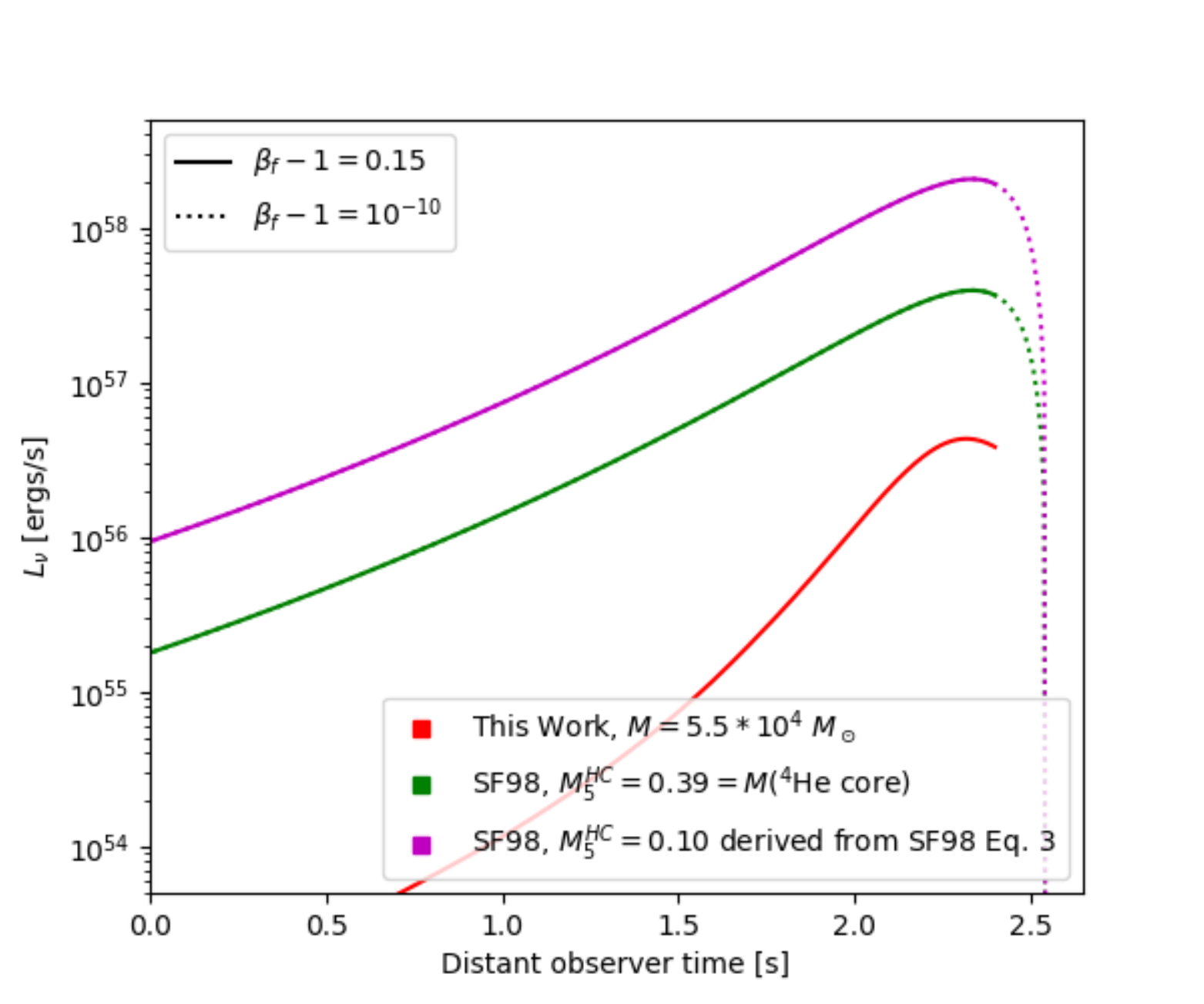}
	
    \caption[Comparison of $L_\nu$ in Fig. \ref{fig:Lt} to Eq. 17 of \citet{shi1998}.]{Comparison of $L_\nu$ in Fig. \ref{fig:Lt} to Eq. 17 of \citet{shi1998}. Since our model does not have a homologous core, we take two reasonable estimates, one large and one small. The solid lines of the result of \citet{shi1998} were chosen with an arbitrary endpoint to match the end of our simulation. Dotted lines show subsequent time evolution.}
    \label{fig:Ltcomp}
\end{figure}

In Fig. \ref{fig:Ltcomp} we compare the distant observer light-curve in Fig. \ref{fig:Lt} with the analytic estimates of \citet{shi1998} for two different values of $M_5^{\rm HC}$. For both values of $M_5^{\rm HC}$, the result of \citet{shi1998} is significantly larger than our result. There are a few reasons for this difference. First, the estimates of \citet{shi1998} are not meant to apply to a situation with significant neutrino trapping, so we would immediately expect the luminosity in our models to be lower than the estimates of \citet{shi1998}. The second reason is that because the inner core of our models does not remain homologous, we have no straightforward way of determining $M_5^{\rm HC}$ which is the primary parameter of the \citet{shi1998} estimate. In Fig. \ref{fig:Ltcomp}, we include two light-curves with different estimates of $M_5^{\rm HC}$. First, we simply assume that $M_5^{\rm HC}$ is the mass of the helium core at dynamical collapse. Next, we use the initial and final values of the core entropy to estimate $M_5^{\rm HC}$ using Eq. 3 of \citet{shi1998}. For each of these light-curves, we assume a minimum value of the final $\beta$ (see \citet{shi1998}) to be $\beta_f - 1 = 0.15$ which was determined by inspection to match the shape of our light-curves. We also include $\beta_f - 1 \ll 1$ in order to show the final behaviour.

Next we discuss the light-curve of a collapsing SMS in comparison with the case of massive star collapse. In Fig. \ref{fig:Lt} the $\Bar{\nu}$ luminosity (blue line) immediately stands out. In CCSN, the neutrino luminosity is dominated by $\nu$ during the neutrinoisation burst \citep{nakazato2013}, but in Fig. \ref{fig:Lt} we can see $L_{\bar{\nu}} > L_\nu$ for almost the entire light-curve. There are two reasons for this. First, the average energy of $\Bar{\nu}$ is higher than that of $\nu$ because of the mass difference between protons and neutrons. Note, however, that because of dilution from pair neutrinos, which have the same energy as each other, this effect is not as prominent as it would otherwise be.

Next, consider the number luminosity of $\nu$ and $\bar{\nu}$. Once again, pair neutrinos contribute equally to $N_\nu,N_{\bar{\nu}}$, but electron/positron capture neutrinos do not. Any change in $X_p$ in the nucleonic region results solely from electron/positron capture and because $X_p = Y_e$, we can use the change in $Y_e$ as a proxy for the number luminosity of different neutrino types. Specifically, if $Y_e$ is increasing, that means positron capture is more frequent than electron capture and so $N_{\bar{\nu}}>N_\nu$. At the neutrino-sphere radius, $Y_e$ is increasing (Fig. \ref{fig:Ye} right panel) for almost the entire simulation, so $N_{\bar{\nu}}>N_\nu$ until the very end of the simulation when $Y_e$ briefly decreases, at which point $N_{\bar{\nu}}<N_\nu$ (Fig. \ref{fig:Lt}).

\begin{table*}
	\centering
	\caption{Summary table for all models. The columns from left to right are mass, core composition at dynamical collapse (see Sec. \ref{progenitor}), the maximum luminosity, total number, and integrated average energy of $\bar{\nu}$ and $\bar{\nu_x}$.}
	\label{tab:models}
	\begin{tabular}{|c|c|c|c|c|c|c|c} 
		\hline
    		M [$10^4$ $\msun$] & Core Composition & Max $L_{ \bar{\nu}}$ [ergs/s] & Max $L_{ \bar{\nu_x}} $ [ergs/s] & Tot $N_{\bar{\nu}}$ & Tot  $N_{\bar{\nu_x}}$ &   $\langle E_{\bar{\nu}} \rangle$ [MeV] &  $\langle E_{\bar{\nu_x}} \rangle$ [MeV] \\
    		\hline \hline
1.0 & OMgSi& 1.07E+56 & 6.12E+55 & 1.35E+60 & 2.37E+59 & 8.20 & 11.37 \\ \hline
2.0 & OMgSi& 2.03E+56 & 1.27E+56 & 3.81E+60 & 8.23E+59 & 7.47 & 10.21 \\ \hline
3.0 & OMgSi& 2.65E+56 & 1.68E+56 & 6.25E+60 & 1.43E+60 & 7.01 & 9.46 \\ \hline
3.5 & HeOMg& 2.98E+56 & 1.91E+56 & 7.10E+60 & 1.71E+60 & 7.01 & 9.41 \\ \hline
4.0 & HeOMg& 3.40E+56 & 2.19E+56 & 9.04E+60 & 2.23E+60 & 6.78 & 9.04 \\ \hline
4.5 & HeOMg& 3.67E+56 & 2.38E+56 & 1.05E+61 & 2.63E+60 & 6.63 & 8.81 \\ \hline
5.0 & HeOMg& 3.99E+56 & 2.58E+56 & 1.24E+61 & 3.11E+60 & 6.46 & 8.55 \\ \hline
5.4 & HeOMg& 4.33E+56 & 2.80E+56 & 1.46E+61 & 3.73E+60 & 6.28 & 8.27 \\ \hline
5.5 & HeOMg& 4.36E+56 & 2.82E+56 & 1.49E+61 & 3.80E+60 & 6.27 & 8.24 \\ \hline
5.6 & HeOMg& 4.24E+56 & 2.73E+56 & 1.40E+61 & 3.57E+60 & 6.33 & 8.33 \\ \hline
6.0 & HeOMg& 4.17E+56 & 2.69E+56 & 1.35E+61 & 3.41E+60 & 6.38 & 8.42 \\ \hline
		\hline
	\end{tabular}
\end{table*}

\subsection{Prospects for neutrino detection}
\label{detection}

\begin{figure}
    \centering
    \includegraphics[width=\columnwidth]{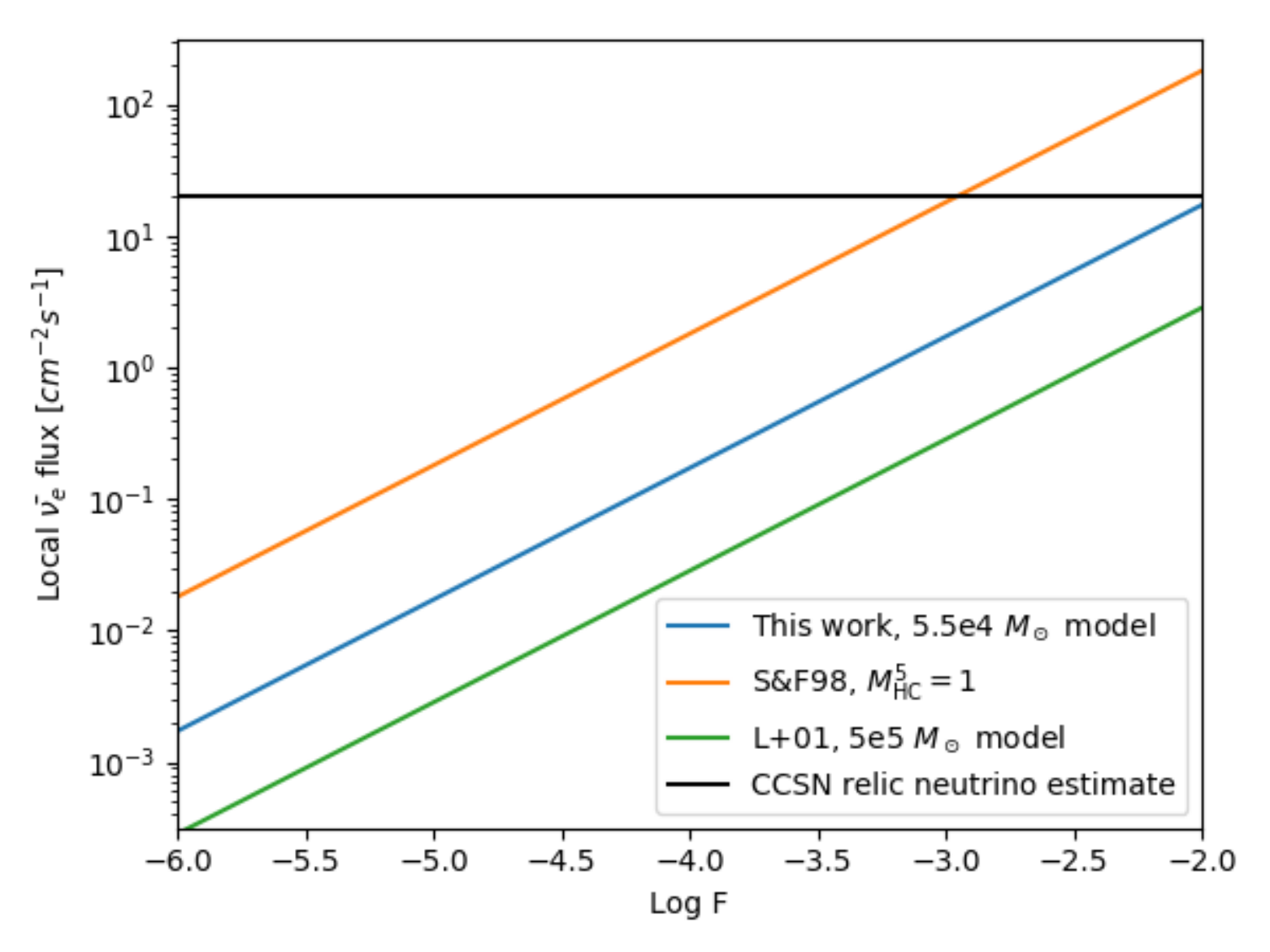}

    \caption{Local anti-neutrino flux using several models as a function of the fraction of baryons which were once in SMSs, F. Even for Log F = -2, the SMS relic neutrino flux is not significantly larger than the CCSN relic neutrino flux. This fact combined with the lower energies of SMS neutrinos in the local universe means a detection of the SMS relic neutrino background is unlikely.}
    \label{fig:local_nu_flux}
\end{figure}

\begin{figure*}
    \centering
    \includegraphics[width=1.0\columnwidth]{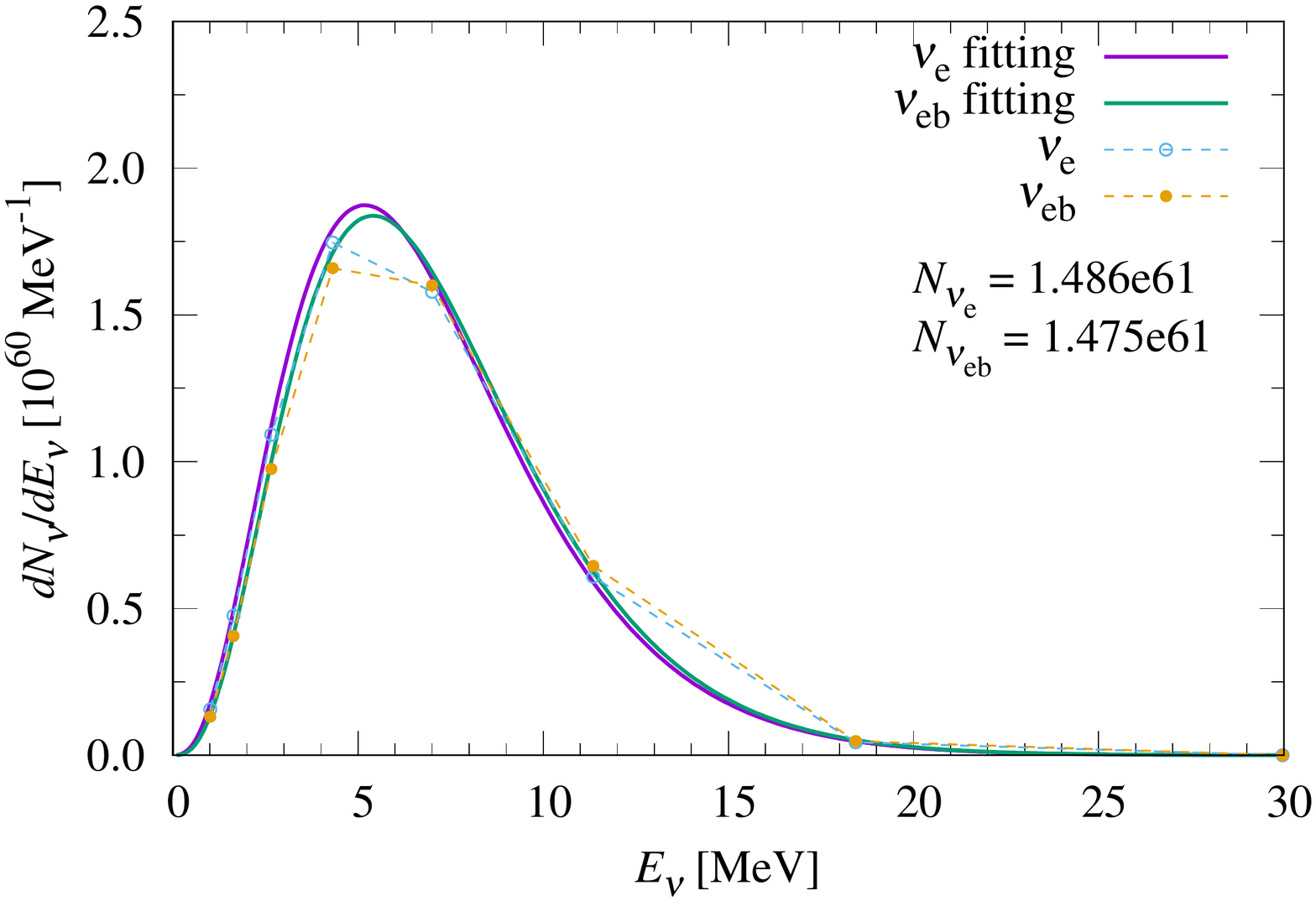}
    \includegraphics[width=1.0\columnwidth]{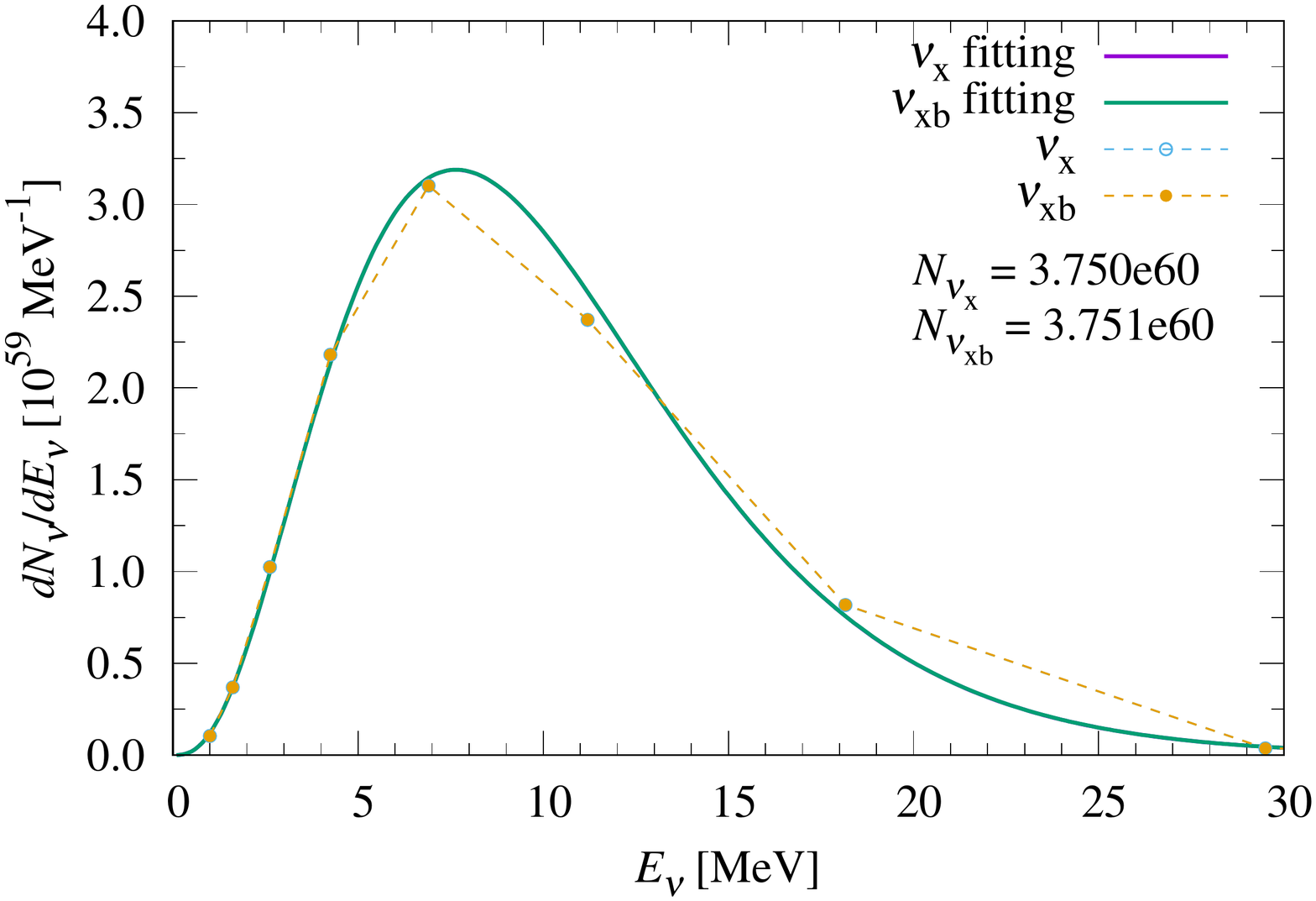}
    
    \caption{Spectra of $\nu,\bar{\nu}$ (left panel) and $\nu_x,\bar{\nu_x}$ (right panel) for the $M = 5.5\times 10^4$ $\msun$ model. The original data is given by the dashed lines where dots show the center of the neutrino energy bins. Solid lines show analytical fits using the fitting formula of \citet{tamborra2012}.}
    \label{fig:spectrum_fit}
\end{figure*}

\begin{figure}
    \centering
    \includegraphics[width=\columnwidth]{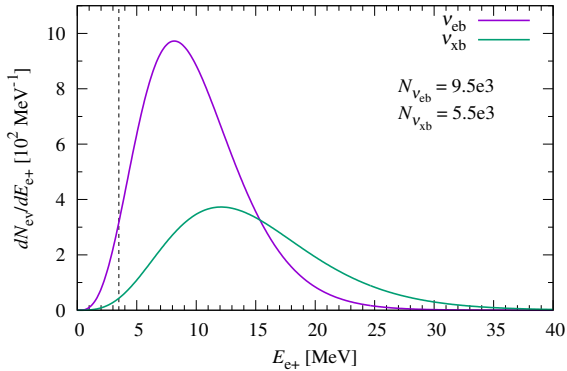}

    \caption{Event rate of a nearby (1 Mpc) supermassive star collapse in SK. This event rate is comparable to a galactic CCSN. When observing neutrinos from a transient event, the detection threshold will be the instrument threshold in SK-IV, around 3.5 MeV.}
    \label{fig:single_event}
\end{figure}

\begin{figure*}
    \centering
    \includegraphics[width=1.0\columnwidth]{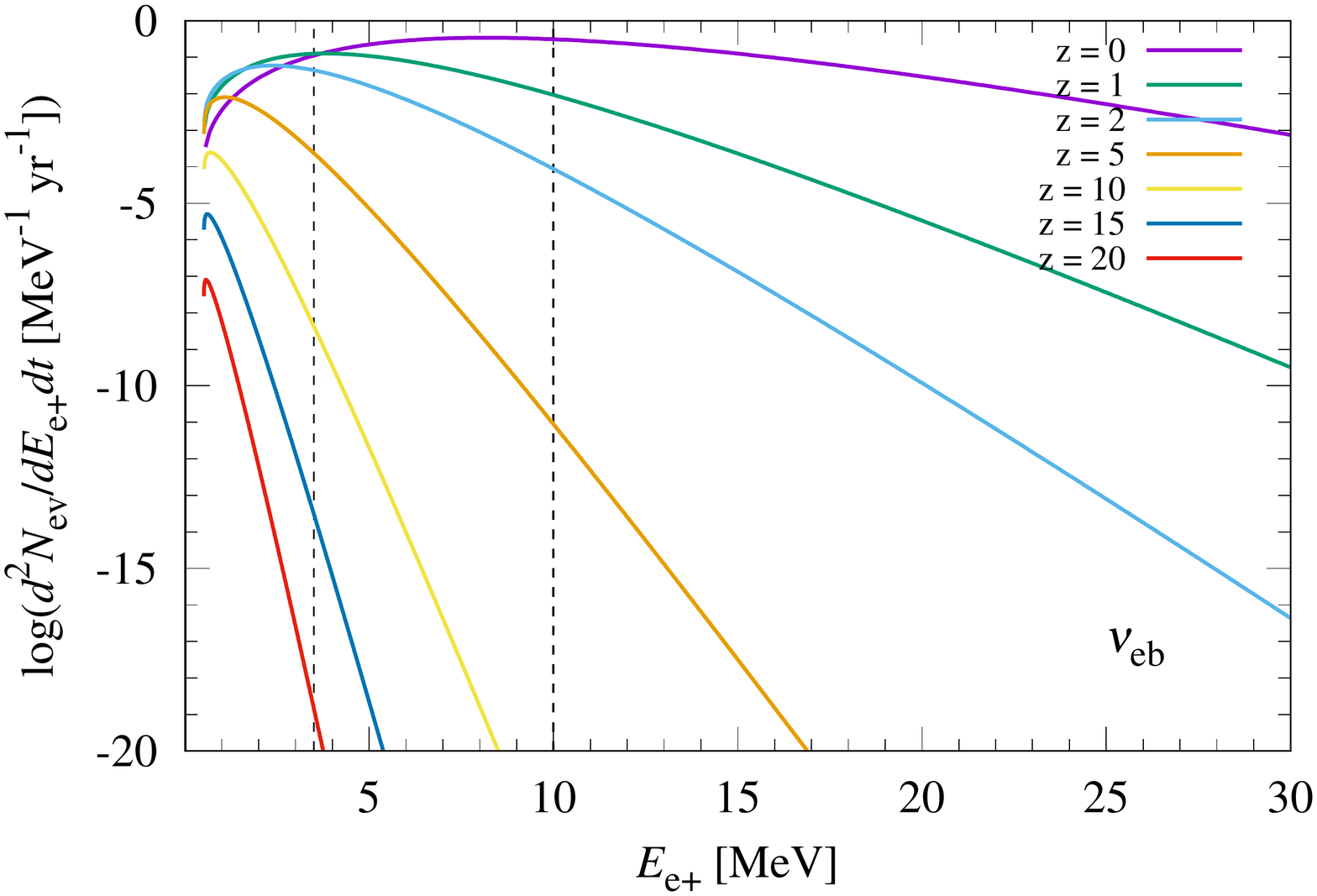}
    \includegraphics[width=1.0\columnwidth]{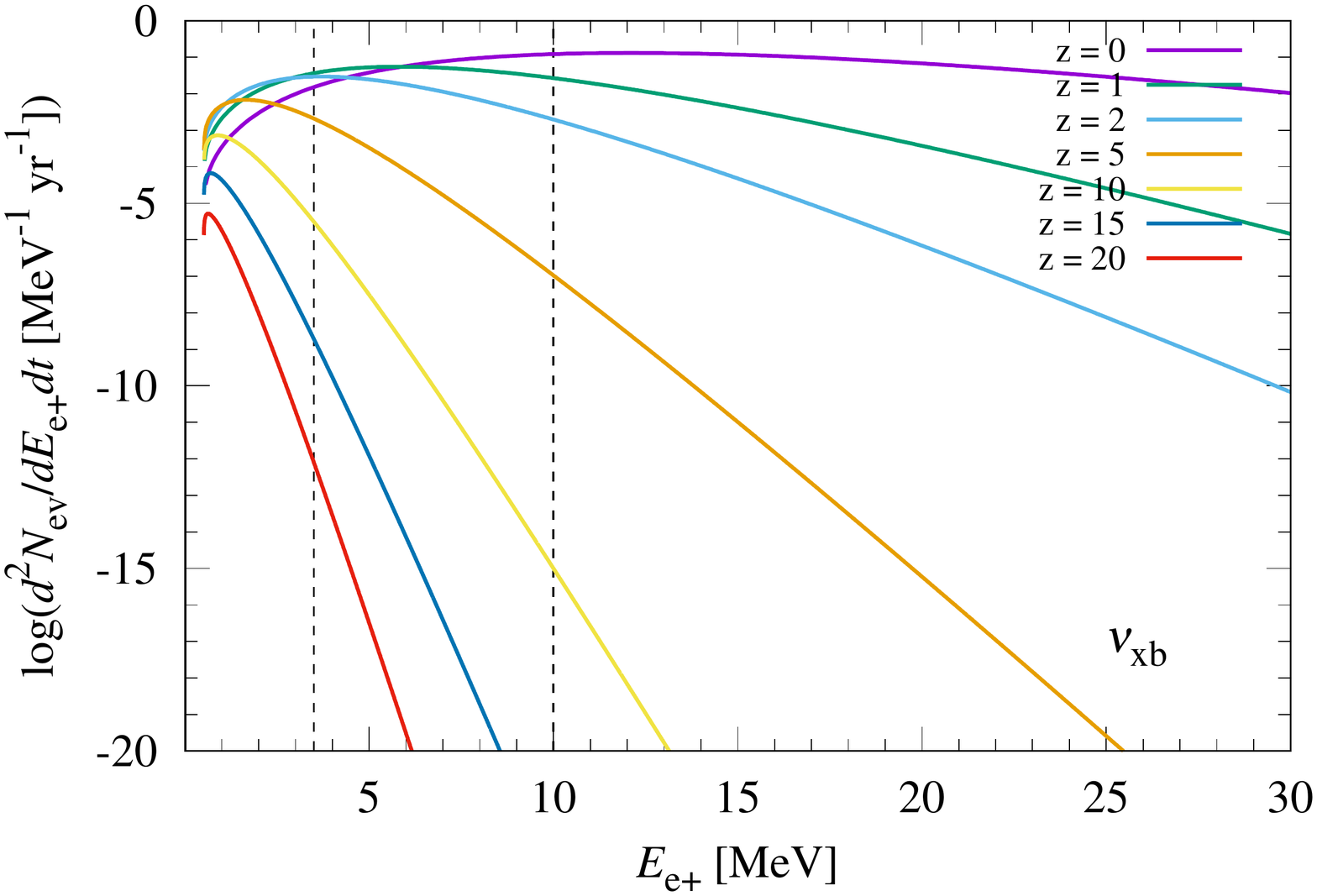}
    \caption{Event rate spectrum of the SMS neutrino background in SK as a function of positron energy for various red-shifts. Dahsed vertical lines are drawn at 3.5 MeV and 10 MeV.}
    \label{fig:SK_E_z}
\end{figure*}

\begin{figure*}
    \centering

    \includegraphics[width=1.0\columnwidth]{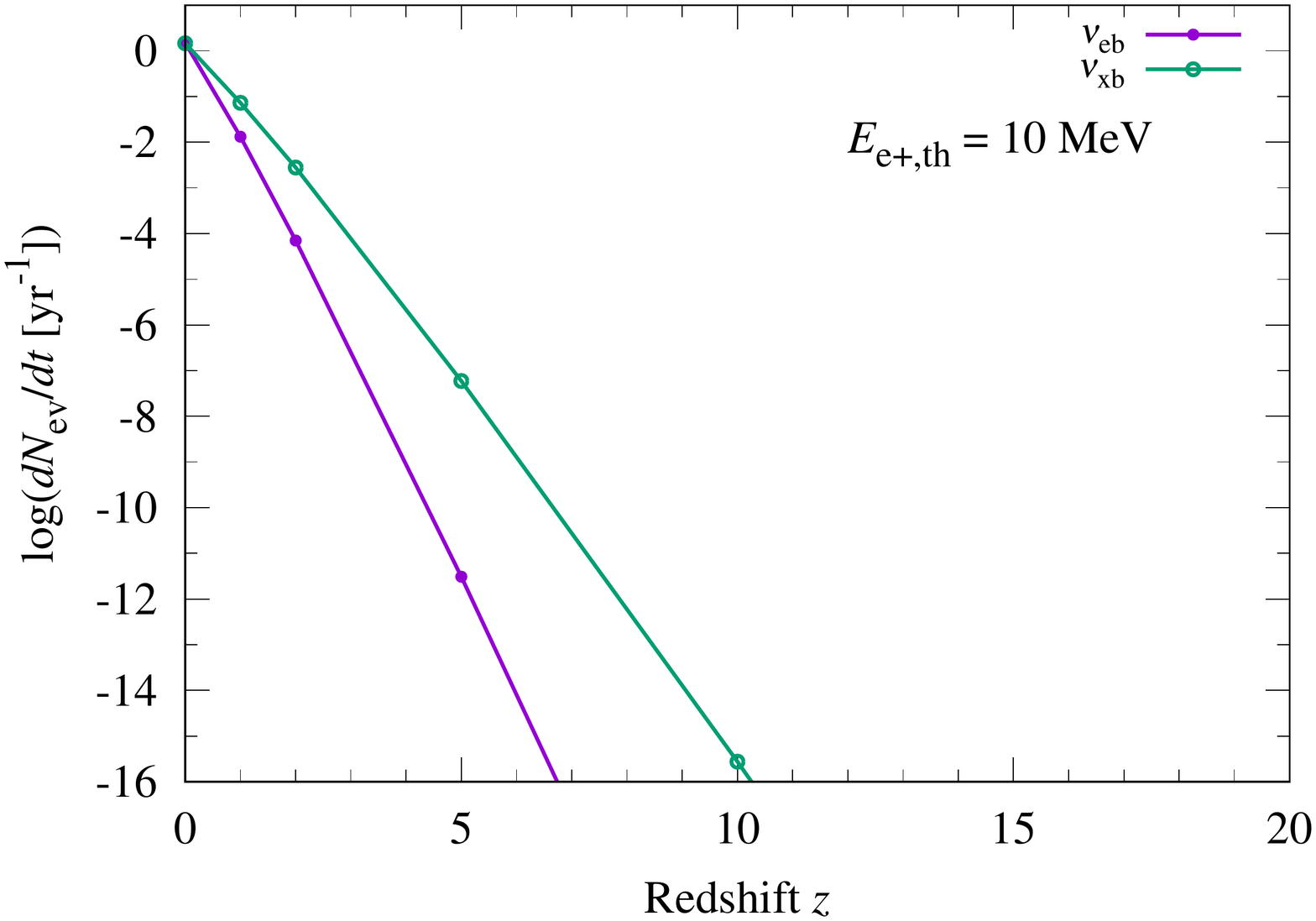}
    \includegraphics[width=1.0\columnwidth]{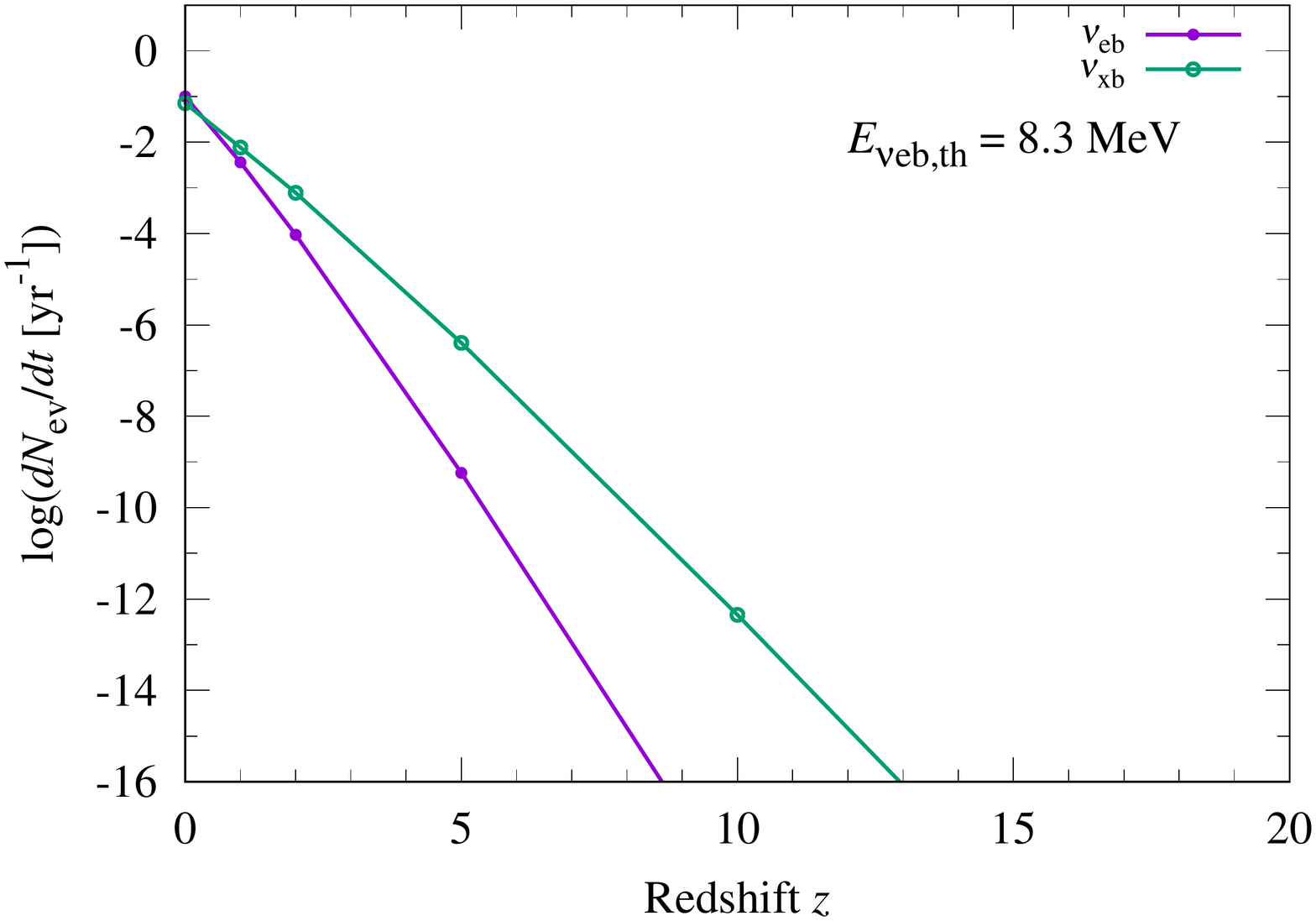}

    \caption{Event rate of the SMS neutrino background in Super Kamiokande with a threshold on the positron kinetic energy of 10 MeV (left panel) and in KamLAND with a threshold on the neutrino energy of 8.3 MeV (right panel) as a function of the redshift at which SMS collapse occurred. The detection of the SMS neutrino background would be very difficult, even taking an optimistic value of $z=2$. The primary reason for this is the lower neutrino energy of SMS neutrinos. }
    \label{fig:SK_z}
\end{figure*}

\begin{figure}
    \centering

    \includegraphics[width=1.0\columnwidth]{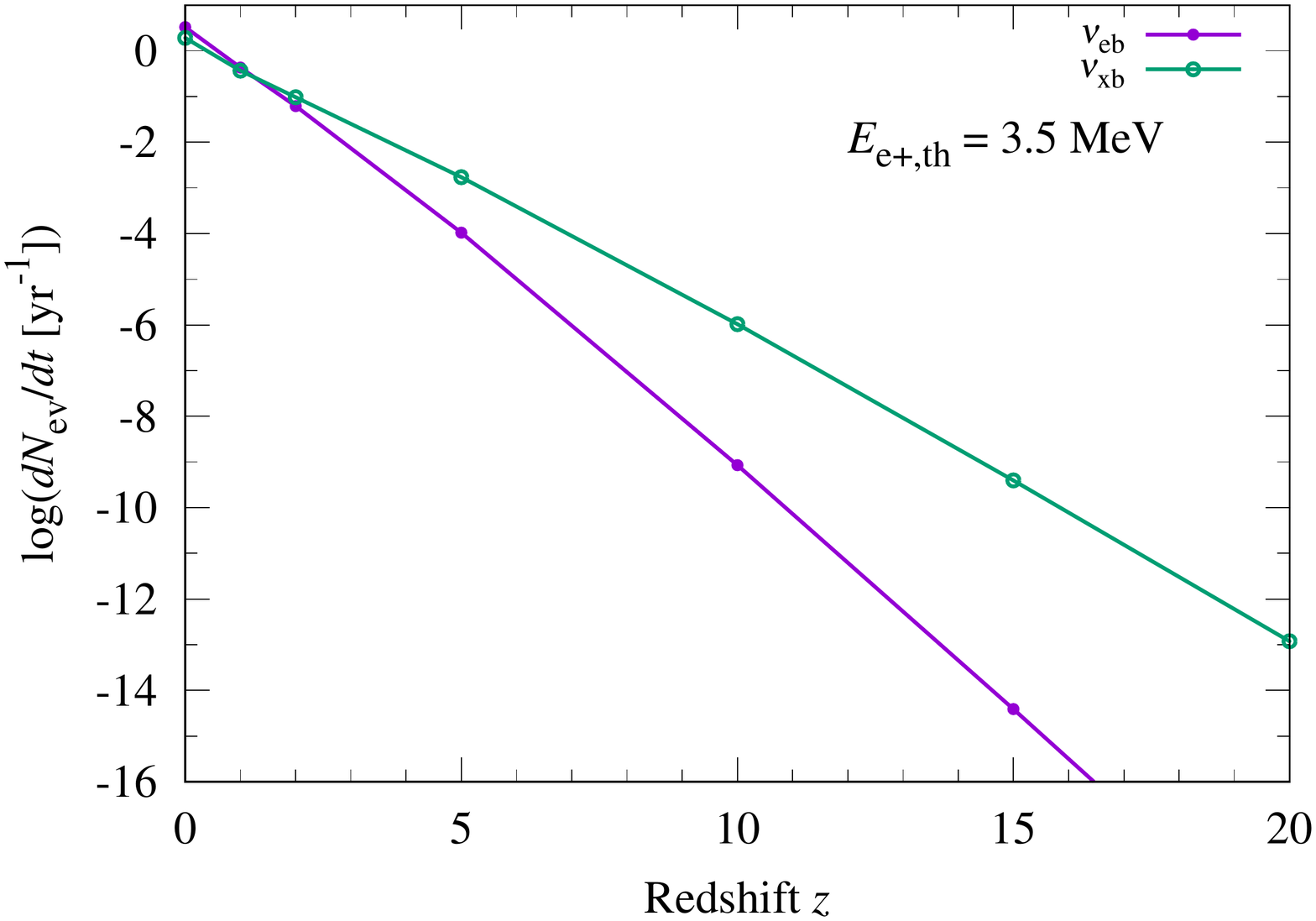}
    
    \caption{Event rate in SK with an observing threshold of 3.5 MeV. If the observing threshold could be reduced to 3.5 MeV, then the situation would become slightly more optimistic, especially once Hyper Kamiokande comes online. However, we should stress that this figure assumed that 1$\%$ of baryons were once in SMSs, and this is an extremely optimistic assumption.}
    \label{fig:SK_3.5}
\end{figure}

It has been suggested that if SMSs are common enough in the early universe, then a strong neutrino background may exist in the local universe \citep{woosley1986}. As in \citet{shi1998}, we will assume that a fraction of baryons in the current universe were once in SMSs and we denote this as $F$. Then we assume that all SMSs have the same mass (in our case, $M=5.5 \times 10^4$ M$\odot$). We integrate the number luminosity in the upper right panel of Fig. \ref{fig:Lt} in order to find the total number of neutrinos per SMS collapse. Then, using the density of baryons in the current universe \citep{PDG2020}
\begin{equation}
    \rho_b = \rho_{\rm crit} \Omega_b = 8.209 \times 10^{9} \;\; \rm \bigg[ \frac{M_\odot}{Mpc^3} \bigg],
\end{equation}
we can estimate the local flux of neutrinos produced by SMS collapse
\begin{equation}
    \phi_\nu = c \frac{N\rho_b}{M} \times F
\end{equation}
as a function of F (Fig. \ref{fig:local_nu_flux}) where c is the speed of light. We can arrive at an extreme upper bound of $F \leq0.01$ by noting that ten percent of baryonic matter resides in galaxies and a SMBH can make up ten percent of a galaxy's baryonic mass. This reasoning assumes that modern SMBHs were once entirely SMSs. In other words, SMBHs are made entirely of black hole mergers with no contribution from accretion. Simulations showing feedback strong enough to disrupt SMBH accretion \citep[e.g.][]{latif2018,chon2021} could support this idea. For the below discussion, we will assume the extremely optimistic $F=0.01$ as in \citealt{shi1998}. 

In order to determine the detectability of this background, we need the neutrino energy spectrum which is obtained from the distribution function:
\begin{equation}
    f = (\hbar c)^3 \frac{\partial^2 n}{\partial x^3 \partial (pc)^3}
\end{equation}
located at the neutrino-sphere for each species. the spectrum is given by
\begin{equation}
    \pdv{n}{E} = \frac{1}{(\hbar c)^3} \int E^2 f c dx^2 dt
\end{equation}
where we have assumed the neutrinos are massless (Fig. \ref{fig:spectrum_fit}, dotted lines). We then fit this spectrum using the formula of \citet{tamborra2012} (Fig. \ref{fig:spectrum_fit}, solid lines). The collapsing SMS spectrum has lower energy than the CCSN relic neutrino spectrum \citep{nakazato2015}, which opens up the possibility of a bi-modal neutrino spectrum with the SMS component at lower energy. 

Note that the spectrum in Fig. \ref{fig:spectrum_fit} may slightly overestimate the distribution of high energy neutrinos. We measure the light-curve at the neutrino-sphere of a neutrino with average energy, so neutrinos with higher than average energy will be inside of their own neutrino-spheres (higher energy neutrinos have a larger neutrino-sphere). We have tested the extent to which the distribution function changes as the neutrinos propagate out of the star --- taking account of gravitational redshift --- and the shift towards lower energy is small but non negligible. In the future, we plan to investigate the neutrino propagation out of the star using the Hernandez Misner metric \citep{hernandez1966}, which is ideally suited for such a situation.

Once the neutrino spectrum is determined, detection rates can be calculated assuming proton absorption of a neutrino $p+\bar{\nu_e} \to n+e^+$:
\begin{equation}
    \dv{N_{\rm event}}{t} = N_p \int \frac{\partial^2 n}{\partial E_\nu \partial t} \sigma_{p+\bar{\nu_e}}(E_\nu) \dv{E_\nu}{E_{e^+}}dE_{e^+}
\end{equation}
where $N_{\rm event}$ is the number of events in the detector, $E_{e^+}$ is the kinetic energy of the positron, $N_p$ the number of protons in the detector, and $\sigma_{p+\bar{\nu_e}}(E_\nu)$ is the cross section \citep{strumia2003}. We assume a detection efficiency of unity. 

As an example, Fig. \ref{fig:single_event} shows the neutrino spectra as a function of positron kinetic energy for a $5.5\times 10^4$ $M_\odot$ SMS collapse located at 1 Mpc detected by SK. The fiducial mass of SK is set to be 22.5 kton. When we assume the positron energy threshold as 3.5 MeV \citep{abe2016}, the event number of $\bar{\nu_e}$ and $\bar{\nu_x}$ for the SMS are $9.5\times 10^3$ and $5.5\times 10^3$ respectively. This event number is comparable to a galactic supernova, because the SMS collapse will produce more neutrinos, but the neutrinos will have a lower energy which makes them harder to detect. Based on the above example, we conclude that detecting a neutrino burst from a collapsing SMS in the high redshift universe is not currently feasible. 

Next we use the calculation shown in Fig. \ref{fig:local_nu_flux} to determine the event rate for the neutrino background from the collapse of $5.5\times 10^4$ $M_\odot$ SMSs at various red-shifts (Fig. \ref{fig:SK_E_z}). Little is known about the formation time of SMSs. \citet{munoz2021} discuss two main possibilities, first that SMS formation aligns with Pop III star formation around redshift 15, and second that SMS collapse aligns with quasar formation around redshift 2. We will consider the latter case which also falls into the extremely optimistic category,

Fig. \ref{fig:SK_z} shows the integral of Fig. \ref{fig:SK_E_z} as a function of redshift for both $\bar{\nu_e}$ and $\bar{\nu_x}$ where we have assumed the threshold of the positron kinetic energy of 10 MeV \citep[e.g.][]{nakazato2015}. We also show the event rate by KamLAND ($N_p= 5.98 \times 10^{31}$) for the neutrino energy threshold of 8.3 MeV \citep[][]{gando2012}. At redshift 2, we might expect one event every thousand years in both Super Kamiokande and KamLAND. The situation improves if the observing threshold in SK could be reduced from 10 MeV to 3.5 MeV, but even in that case we would not see more than 1 event every ten years. Since we have used two extremely optimistic assumptions to reach this number, the natural conclusion is that a detection of the SMS neutrino background is extremely unlikely. 

Two possibilities exist to alter this conclusion. The first is that the neutrino number from each collapse event is underestimated. This could occur if SMSs are more massive than $5.5 \times 10^4$ $\msun$ (and less massive than $5 \times 10^5$ $\msun$ judging by the results of \citealt{linke2001} in Fig. \ref{fig:local_nu_flux}) and have a higher neutrino luminosity, or if there is significant neutrino emission during the accretion of the core and the envelope (see e.g. \citealt{mclaughlin2007}). We plan to investigate this in future work. The second possibility is if neutrino detectors were to improve, either by increasing the detector mass or decreasing the detection threshold.

\section{Discussion}
\label{discussion}

We will now discuss the results from these simulations relative to \citet{shi1998} and \citet{linke2001}. When it comes to comparing our neutrino light-curve to that of \citet{shi1998}, there are two main differences, namely the scale and the shape of the light-curve. In the previous section we discussed how the results of \citet{shi1998} are intended to apply to SMSs which have a homologous core and which do not exhibit neutrino trapping. Since our models do have trapping and do not have a homologous core, the comparison is challenging. However, broadly speaking we can say that the luminosities predicted by \citet{shi1998} are larger than our result by one to two orders of magnitude. We cannot compare directly to the other numerical studies which used more massive progenitors, but those studies can be compared to the formula of \citet{shi1998} and to each other. Specifically, \citet{linke2001} found that their neutrino luminosity was $3\%$ that of \citet{shi1998} and half that of \citet{woosley1986}. We note that both \citet{shi1998} and \citet{linke2001} include neutrino emission via pair production, but they do not include other neutrino reactions (\citet{linke2001} also includes photo-neutrino emission and plasmon decay, but for low mass models those reactions are negligible, see Fig. 5 of \citealt{linke2001}). In our mass range, these missing neutrino reactions are important. On the emission side, neutrinos from electron/positron capture are almost as common as pair neutrinos (Fig. \ref{fig:emiss}). On the trapping side, we include pair annihilation, electron scattering, and nucleon absorption, all of which are significant sources of opacity in our simulations (Fig. \ref{fig:mfp}). 

Besides the core structure and neutrino trapping, there is an additional possible cause for the different luminosities. Specifically, neutrino trapping causes the gravitational mass in the inner core to increase and causes a smaller apparent horizon to form for lower $T$ than was found in earlier works. \citet{linke2001} found that the initial apparent horizon enclosed $25\%$ of the total stellar mass and \citet{woosley1986} found $20\%$ of the total stellar mass whereas we find values in the range of $1-2\%$ of the total stellar mass.

\citet{linke2001} claim that after an apparent horizon forms, the temperature will not increase further. We are not sure if this claim is true in a neutrino trapping situation, but if it is, it could explain the lower $T$ and hence lower $L$ of our simulation compared to \citet{shi1998}. Indeed, \citet{li2018} pointed out that the exact physics around the apparent horizon may have a large effect on the outcome and may explain the discrepancy between \citet{shi1998} and \citet{linke2001}. Finally, the shapes of our light-curve and the light-curve of \citet{shi1998} are fairly consistent. The only major difference is the steeper slope of our light-curve which is likely due to the fully GR nature of our code.

Previous works considered that collapsing SMSs could be parameterized by a single parameter, the mass of the homologous core, e.g. because of the dependence of the critical density solely on mass \citep{shi1998,linke2001}. In this study, we find that the central entropy more accurately parameterizes our models, because the substantial nuclear burning between the GR instability and dynamical collapse does not depend on mass in a straightforward way \citep{nagele2020}. Once dynamical collapse has started, the entropy hierarchy remains fixed. 

\citet{shi1998} argue and \citet{linke2001} confirm that the neutrino luminosity scales inversely with mass, because more massive stars form black holes with lower temperatures. We find that $T_{\rm c}$ at apparent horizon formation (Fig. \ref{fig:scomp} first lower panel) scales inversely with entropy. This means that it will also scale roughly inversely with mass. However, where \citet{shi1998} and \citet{linke2001} found that this meant $L_\nu$ would also scale inversely as a function of mass, we find the opposite (Fig. \ref{fig:scomp} third upper panel). Essentially this is because while final $T$ may decrease as a function of entropy, $T$ at fixed $\rho$ increases as a function of entropy, and $T$ at fixed $\rho$ is what determines the neutrino trapping and thus emission. As we increase entropy, we would expect the neutrino-sphere radius to also increase. However, as entropy increases, the apparent horizon size increases, and Fig. \ref{fig:scomp} second lower panel shows that the final apparent horizon is getting closer to the neutrino-sphere radius. Thus, for high enough entropy, we would expect the final apparent horizon to fall outside the neutrino-sphere, and for even higher entropy, the neutrino-sphere may fall inside the apparent horizon at all times. In the former case, the neutrino energy density would be  decreased (relative to other models) when the apparent horizon eclipsed the neutrino-sphere, while in the latter case, we expect the neutrino energy density to be greatly decreased. Investigating higher mass models to find exactly where each of these events occur will be the topic of future work.

We have calculated event rates in SK and KamLAND for neutrinos from SMS collapse. We find that a nearby SMS collapse would be detectable, but that the neutrino background from SMS collapse is not detectable for any reasonable assumptions about the redshift of SMSs. This is in agreement with the results of \citet{munoz2021} who considered a similar question for detectors using coherent neutrino scattering.

As well as the sources of error discussed in Sec. \ref{results}, we should consider sources of error directly related to the calculation of the neutrino light-curve. Firstly, since the nuRADHYD code terminates soon after the apparent horizon formation, the assertion that the neutrino luminosity will continue to decrease past the end of the current light-curve may not be valid. In particular, the star has a significant amount of matter outside the apparent horizon which will accrete onto the black hole, and this could lead to further neutrino emission \citep[e.g. Figs. 9, 13 of][]{sekiguchi2011}. Next, we must consider trapping outside the neutrino-sphere. This should not be an extremely large effect, but it is present and would further reduce the number luminosity of high energy neutrinos. In future work we intend to run a version of nuRADHYD using the Hernandez Misner scheme which avoids the event horizon. Such a simulation would be able to track the neutrinos as they propagated out of the star and accurately determine the amount of neutrino trapping. Finally, we should also consider multidimensional effects, though these effects may not change the total neutrino energy as evidenced by the agreement of neutrino energy in \citet{linke2001} and \citet{montero2012} to within ten percent. Multidimensional effects can change the angular temperature distribution \citep{li2018} and in addition the apparent horizon formation may be anisotropic, both of which could effect the neutrino light-curve. Furthermore, multidimensional effects will be important for the accretion of material onto the black hole \citep{sekiguchi2011}.

\section*{Acknowledgements}
This study was supported in part by the Grant-in-Aid for the Scientific Research of Japan Society for the Promotion of Science (JSPS, Nos. JP17K05380, JP19K03837, JP20H01905, JP20H00158, JP21H01123) and by Grant-in-Aid for Scientific Research on Innovative areas (JP17H06357, JP17H06365, JP20H05249) from the Ministry of Education, Culture, Sports, Science and Technology (MEXT), Japan. For providing high performance computing resources, YITP, Kyoto University is acknowledged. K.S. would also like to acknowledge computing resources at KEK and RCNP Osaka University. T.Y. thanks Koji Ishidoshiro for discussing neutrino detection by KamLAND.



\bibliographystyle{mnras}
\bibliography{bib}




\section*{Appendix A: Calculation of Neutrino-sphere}

In this appendix, we will calculate the electron type neutrino-sphere using the temperature, density, and average neutrino energy and compare the result to that recorded by the simulation. First, let's consider the nucleon neutrino absorption reactions which are the primary source of opacity in a CCSN. The reaction $(n + \nu \;\to\; e^- + p)$ has a cross section
    \begin{equation}
        \sigma = \sigma_0 \frac{(1+3a^2)C_V^2}{4} \bigg(\frac{E_\nu}{m_ec^2}\bigg)^2 \bigg(1+\frac{Q}{E_\nu} \bigg) \bigg[\bigg(1+\frac{Q}{E_\nu}\bigg)^2-\bigg(\frac{m_ec^2}{E_\nu}\bigg)^2\bigg]^{1/2} I(E_\nu + Q)
    \end{equation}
where $\sigma_0=1.76 \times 10^{-44}$ cm$^2$, $a$ and $C_V$ are coupling constants, Q is the exothermic energy, $m_e$ is the electron mass, and $I$ is a statistical inhibition factor \citep{shapiro1983}.

The primary dependence of this cross section in a non degenerate regime is $\sigma \propto E_\nu^2$ and we will assume that the average energy scales with temperature in the trapping regions $\langle E_\nu \rangle^2 \propto T^2$, meaning 

    \begin{equation}
        \lambda_{MFP} (\langle E_\nu \rangle) = \frac{1}{n \: \sigma (\langle E_\nu \rangle)} \propto \frac{1}{\rho T^2}.
    \end{equation}

For a high temperature, low density environment, the neutrino mean free path (MFP) will decrease. As a rough approximation, the neutrino-sphere radius scales as $\lambda_{MFP} (\langle E_\nu \rangle)^{-1}$, so that the neutrino-sphere of the collapsing SMS will scale as $R_{\nu} \propto \rho T^2 $. At fixed density, 
\begin{equation}
    \frac{T_{\rm SMS}}{T_{\rm CCSN}} \sim 10 \Rightarrow \frac{R_{\nu, SMS}}{R_{ \nu, CCSN}} \sim 100
    \label{eq:nu_demo}
\end{equation}
so despite the low density of the SMS, the neutrino-sphere is significantly larger than in the case of CCSN. Typical densities at the neutrino-sphere radius are $\rho_\nu \sim 10^7$ g/cc at the neutrino-sphere formation and  $\rho_\nu \sim 10^8$ g/cc at the final time step of the simulation. 

Above we have shown that a neutrino-sphere can form in SMSs only by considering the differences in temperature and density between SMSs and CCSN. Next, we will discuss the electron scattering reaction $(e^- + \nu \;\to\; e^- + \nu)$ which is not important for CCSN because of electron degeneracy. In a collapsing SMS, degeneracy is negligible because the low density and the high temperature means pairs are plentiful. Thus there are enough electrons to cause significant opacity for the neutrinos. The cross section can be approximated using \citep{bahcall1964}:

\begin{equation}
    \sigma = 1.92\; \sigma_0 \bigg(\frac{kT}{m_ec^2}\bigg) \bigg(\frac{E_\nu}{m_ec^2}\bigg) \bigg(\frac{1}{2}+\frac{1}{6}\bigg),
\end{equation}
where the factor of 1/2 accounts for $e^- + \nu$ and the factor of 1/6 accounts for $e^+ + \nu$. Note that this cross section also has $T^2$ dependence in the thermalized neutrino region. For all of our models, the electron scattering reaction is the primary reaction for determining the location of the neutrino-sphere (Fig. \ref{fig:mfp}). However, we also note that ignoring the electron scattering reaction would not decrease the neutrino-sphere radius significantly.

Next, consider nucleon scattering, $(n + \nu \;\to\; n + \nu)$ which has a cross section,

\begin{equation}
     \sigma = \frac{\sigma_0}{4}  \bigg(\frac{E_\nu}{m_ec^2}\bigg)^2   
\end{equation}
\citep{shapiro1983}. This also has $T^2$ dependence in the trapping region though because the nucleon number density is lower than the charged lepton number density, this reaction is not as important as the above two reactions.

Finally, pair neutrino annihilation ($\nu + \Bar{\nu} \to e^- +e^+$)  is one of the primary sources of opacity in the center of the star at late times, but does not contribute to the opacity near the neutrino-sphere.

It is a useful exercise (and sanity check) to compare the above estimates to the simulation. The treatment in nuRADHYD will be more accurate than the above expressions because it includes a grid in neutrino energy space and the full integration over lepton and baryon momentum space. The simulation records the mean free path of a neutrino in a specific energy bin for each radial mesh and neutrino reaction. Fig. \ref{fig:mfp}. shows the mean free paths of neutrinos with energy $12.9$ MeV from the simulation (lines) and calculated from cross sections (symbols). $12.9$ MeV is the closest energy bin to the average energy around the neutrino-sphere radius, so the mean free path in Fig. \ref{fig:mfp} is approximately the mean free path of a neutrino with average energy.

In Fig. \ref{fig:mfp}, we can see that the electron scattering expression is fairly accurate, while the nucleon absorption and nucleon scattering mean free paths are both slight underestimates. Still, the total mean free path agrees remarkably well with the simulation and a neutrino-sphere at $R_\nu \approx 3 \times 10^9$ cm can be clearly observed in the right panel. When determining the neutrino-sphere in the rest of this paper, we will use the usual condition on the optical depth: 
\begin{equation}
    R_\nu = r \;|\; \tau(r,\langle E \rangle) = 2/3.
\end{equation}
This condition agrees well with the result from the above discussion of mean free paths.

\section*{Appendix B: Neutrino thermalization}

\begin{figure}
    \centering
    \includegraphics[width=\columnwidth]{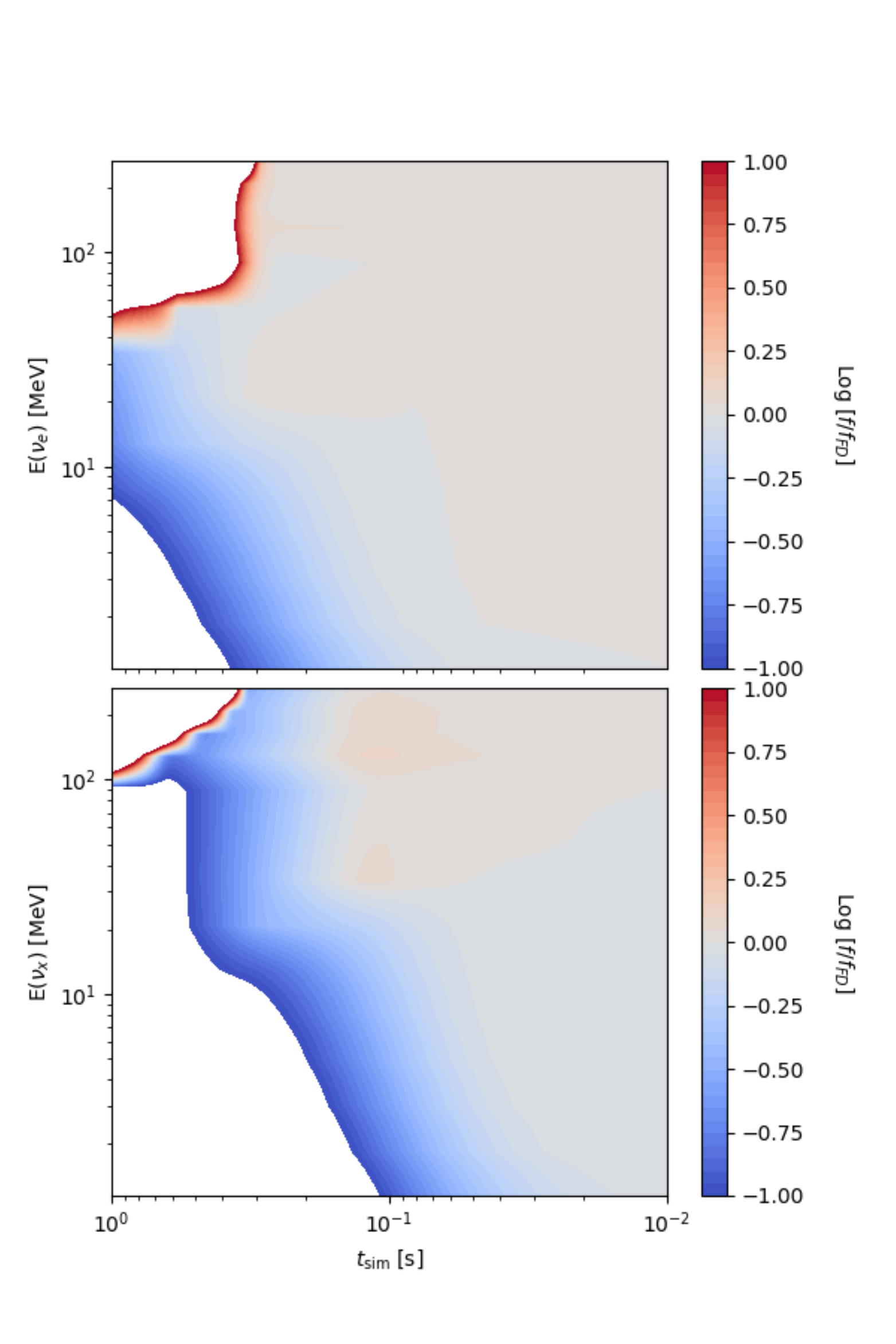}

    \caption{Ratio of the distribution function to the Fermi-Dirac distribution for the central mesh as a function of time and neutrino energy for $\nu_e$ (upper panel) and $\nu_x$ (lower panel). Blue areas, such as the bottom left, show regions where the simulation distribution is below the FD distribution, while red, such as the top central, show regions where the distribution is above FD. White areas are not covered by the range of the colorbar because of numerical noise (upper left) or large physical mismatch (lower left). A neutrino species is said to be thermalized when $f=f_{FD}$ (grey) for all energy bins. }
    \label{fig:B2}
\end{figure}

\begin{figure}
    \centering
    \includegraphics[width=\columnwidth]{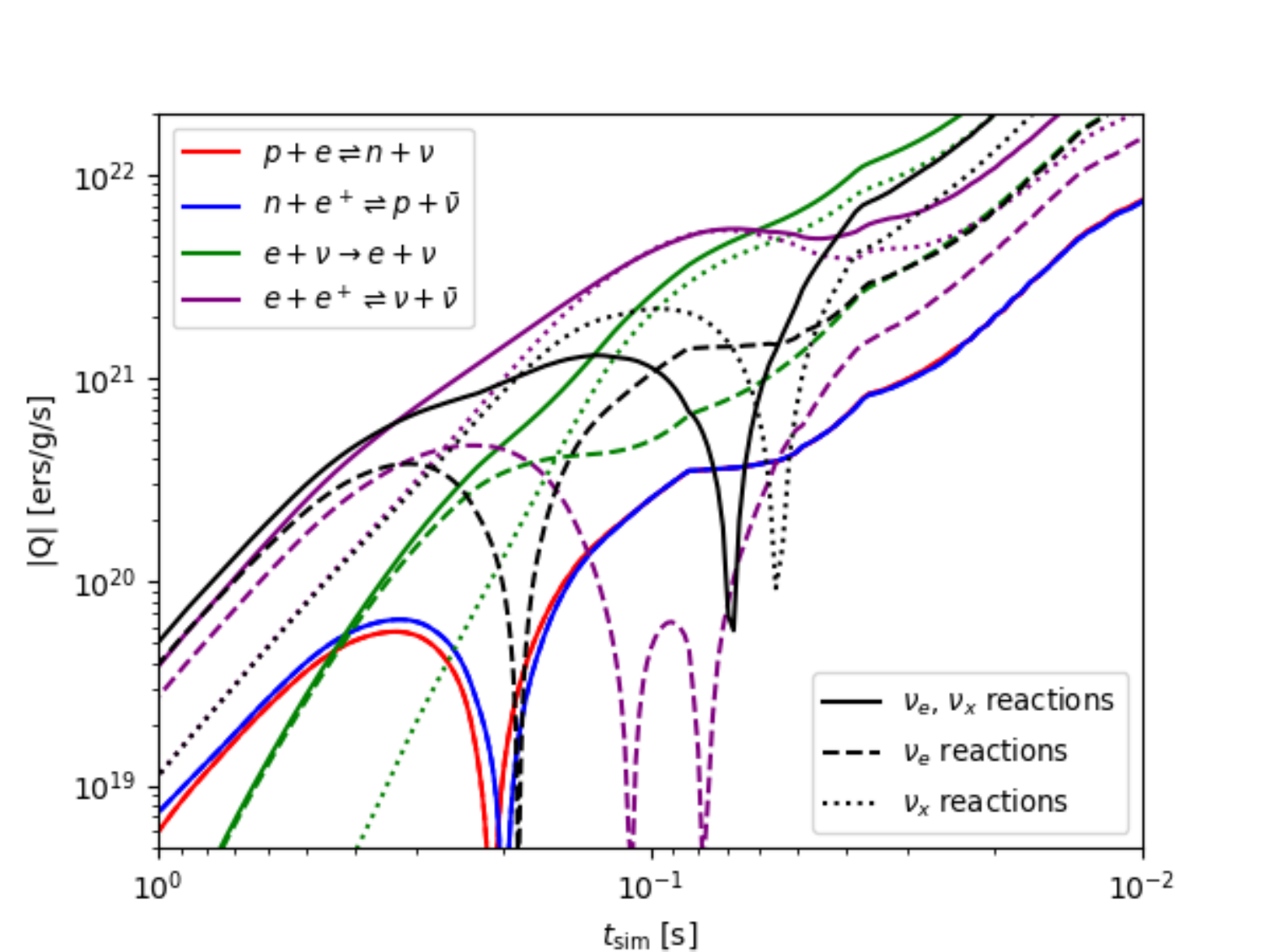}

    \caption{Decomposition of the emissivity from Fig. \ref{fig:emiss} into reactions involving e-type neutrinos (dashed lines) and x-type neutrinos (dotted lines). The nuRADHYD code does not include muons so nucleon capture of $\mu, \bar{\mu}$ and pair annihilation of $\mu, \bar{\mu}$ do not contribute to these reaction rates. All reaction rates are initially negative except for electron scattering. Several reaction rates change sign (twice for pair processes on electron type neutrinos) which can be seen by the downwards spikes. Note that the changes in sign of the total rates correspond to thermalization of e-type and x-type neutrinos and e-type thermalization occurs first. The change in sign of the overall reaction rate occurs in between the change in sign of the two flavors.     }
    \label{fig:B1}
\end{figure}

The increase in entropy in Fig. \ref{fig:Mrs} coincides with the thermalization of neutrinos with matter and photons. This phenomenon can be measured by determining when the neutrino distribution function matches the Fermi-Dirac distribution, which can be calculated in terms of the neutrino chemical potential and temperature:

\begin{equation}
    f_{FD} = \bigg(1+e^{(\epsilon - \mu_\nu)/T}\bigg)^{-1}.
\end{equation}

The ratio $f/f_{FD}$ for $\nu_e$ and $\nu_x$ is shown in Fig. \ref{fig:B2}. The simulation distribution function first matches (or even slightly exceeds) the Fermi Dirac distribution for high energy (>20 MeV) pair neutrinos. These pair neutrinos then either annihilate with other high energy neutrinos or down scatter on electrons and positrons. The combination of these two processes heats the matter and photons.

Fig. \ref{fig:B2} show that the different neutrino flavors thermalize at different times and Fig. \ref{fig:B1} shows the flavor decomposition of emissivity Q from Fig. \ref{fig:emiss}. In the first part of the simulation, the net neutrino emissivity is a cooling reaction. At $t_{\rm sim} = 0.2$, the emissivity for $e$-type neutrinos becomes positive and shortly thereafter, the pair emissivity for $e$-type neutrinos becomes briefly positive. However, by this point, $x$-type neutrino reactions dominate the energy change and the overall emissivity stays negative until $t_{\rm sim} = 0.06$ when the $x$-type neutrinos begin to thermalize, and this can be seen by the decrease in the $x$-type pair emissivity (purple dotted line). Note that the sign changes in Fig. \ref{fig:B1} occur as $f/f_{FD} \to 1$ in Fig. \ref{fig:B2}.

One possible source of error in our calculation --- though it should not effect the neutrino light-curve --- particularly for $T>10$ MeV is that the nuRADHYD code does not include muons, which have been shown to soften the EOS in CCSNe \citep{bollig2017}. As an example, at the time of thermalization of $\nu_x$, $n_\mu / n_e \sim 10^{-5}$ and it is conceivable that the inclusion of muon reactions could affect the thermalization. More detailed calculations would be needed to determine if muon pair annihilation could be a significant source of $\nu_\mu$ compared to $\nu +\bar{\nu} \to \nu_\mu +\bar{\nu_\mu}$ and $e^- + e^+ \to \nu_\mu +\bar{\nu_\mu}$ (for a discussion of the relative strength of these two reactions, see \citealt{buras2003}).

\bsp	
\label{lastpage}
\end{document}